\documentclass[aps,pre,twocolumn,superscriptaddress,amssymb]{revtex4-2}
\usepackage[pdftex]{graphicx}
\usepackage{amsmath,amsthm,amssymb,cases}
\begin{document}

\title{Higher-order tensor renormalization group study of the $J_1$-$J_2$ Ising model on a square lattice}
\author{Kota Yoshiyama}
\affiliation{
 Graduate School of Arts and Sciences,
 The University of Tokyo,
 Komaba, Meguro-ku, Tokyo 153-8902, Japan
}

\author{Koji Hukushima}
\affiliation{
 Graduate School of Arts and Sciences,
 The University of Tokyo,
 Komaba, Meguro-ku, Tokyo 153-8902, Japan
}
\affiliation{
Komaba Institute for Science, The
University of Tokyo, 3-8-1 Komaba, Meguro-ku, Tokyo 153-8902, Japan
}
\date{\today}

\begin{abstract}
    Phase transitions of the $J_1$-$J_2$ Ising model on a square lattice are studied using the higher-order tensor renormalization group(HOTRG) method. This system involves a competition between the ferromagnetic interaction $J_1$ and antiferromagnetic interaction $J_2$. Furthermore, weak first-order and second-order transitions are observed near the ratio $g=J_2/|J_1|=1/2$. Our results (based on HOTRG calculations for significantly larger sizes) indicate that the region of the first-order transition is marginally narrower than that in previous studies. Moreover, the universality class of the second-order transition connected to the transition line is not necessarily fully consistent with the Ashkin--Teller class considered earlier.
\end{abstract}

\maketitle

\section{Introduction}
Frustrated magnetic systems have been studied for a long time since the concept of frustration in magnetic systems was proposed\cite{Toulouse1977,Villain1977} (see for example the review\cite{Diep2013} and references therein).
One of the simplest models is the $J_1$-$J_2$ Ising model on a two-dimensional square lattice with ferromagnetic interactions between nearest neighbors and antiferromagnetic interactions between next-nearest neighbors. Although the model is highly simple, it remains unclear in many respects and has been studied extensively till recently\cite{Jin2012,Jin2013,Kalz2012,Li2021,Hu2021}.

The model involves a parameter $g$, which is the ratio of the amplitudes of antiferromagnetic and ferromagnetic interactions. It determines the characteristics of phase transitions. In this model, $g=1/2$ is a special point, where the ground state transforms from a fully ferromagnetic state to a stripe-ordered state. As the temperature is lower, the phase transition occurs to the ferromagnetic phase for $g<1/2$ and to the phase with a stripe order for $g>1/2$. Many studies have asserted that the phase transition for $g<1/2$ belongs to the Ising universality class, although it is not a complete conclusion\cite{Hu2021}.

The phase transition for $g>1/2$ has also been studied by many methods including the variational mean-field theory\cite{Jin2013}, and numerical methods such as the Monte Carlo method\cite{Jin2012,Jin2013,Kalz2012} and transfer matrix methods\cite{Jin2013,Hu2021}. An issue being discussed is whether this model shows a first-order or second-order transition in the vicinity of $g=1/2$.
The first-order transition in this system, if it exists, has been indicated to be highly weak. Thus it is in general difficult to determine whether it is a first-order or second-order transition. Therefore, a few previous studies have made different assertions on the existence or non-existence of the region of the first-order transition, and on the width of the region of the first-order transition.

Because systems exhibiting a weak first-order transition have a finite but significantly large correlation length at the transition temperature, it is necessary to investigate systems with sizes larger than the correlation length to clarify the characteristic of phase transition by numerical simulations.  Tensor renormalization group (TRG) methods\cite{Levin2007} have attracted attention recently. These are potential numerical methods that can be computed to sizes significantly larger than those achieved by existing methods. In this method, the system is represented by a tensor network (TN), and a renormalization calculation is performed to approximate its contraction. Under a certain assumption, one can compute the free energy of the system with a computational complexity with the logarithm of the system size. In this study, we use the higher-order tensor renormalization group (HOTRG)\cite{Xie2012} method. It is a variant of the TRG. The method has the advantage of calculating higher-order derivatives of the free energy. It has been demonstrated that it clearly distinguishes between first- and second-order phase transitions, by applying it to the two- to six-state Potts model\cite{Morita2019}. Utilizing these advantages, we perform statistical mechanics calculations for the $J_1$-$J_2$ Ising model of large sizes to obtain the details of its phase diagram.

Our results of the HOTRG calculations indicate that the first-order transition exists in a finite region of the parameter $g$. However, the region is narrower than concluded in the previous MCMC study\cite{Jin2012}. We also verify that the universality class of the second-order transition connected to the first-order transition line is consistent with the Ashkin--Teller (AT) weak universality (indicated in the previous study). Under the weak universality, critical exponents depend explicitly on the parameter $g$. However, their ratios remain constant. In contrast, our results also indicate that the critical exponent $\nu$ of the correlation length can adopt smaller values beyond the lower bound of the range varying in the AT universality class. This implies that the correspondence between the $J_1$-$J_2$ Ising model and AT model cannot be naively accepted.

This paper is organized as follows. First, the model investigated in this study is explained in Sec.~\ref{sec:model}. Next, the numerical method used, HOTRG, is described in Sec.~\ref{sec:methods}. The results obtained by modifying the model parameters are explained in Sec.~\ref{sec:results}. Sec.~\ref{sec:discussion} presents the discussions and a summary of this work. The Appendix presents certain discussions on numerical validations.

\section{Model}
\label{sec:model}
The $J_1$-$J_2$ Ising model has a ferromagnetic interaction between the nearest-neighbor spins and an antiferromagnetic interaction between the next-nearest neighbor spins. The Hamiltonian is expressed by
\begin{equation}
    H = J_1\sum_{\langle i,j\rangle} \sigma_i \sigma_j +J_2\sum_{\langle\langle i,j\rangle\rangle} \sigma_i \sigma_j,
    \label{eqn:Hamiltonian}
\end{equation}
where $\sum_{\langle i,j\rangle}$ and $\sum_{\langle \langle i,j\rangle\rangle}$ represent the sums over the nearest-neighbor and next-neighbor sites, respectively. Meanwhile, $J_1$ and $J_2$ are the ferromagnetic and antiferromagnetic exchange interaction energies. These satisfy $J_1<0$ and $J_2>0$.
In the following, $J_1$ is taken as the unit of energy. The phase transition of this model is discussed with different values of the parameter $g\equiv-\frac{J_2}{J_1}$, which adopts a positive value.

\begin{figure}[ht]
 \centering
 \includegraphics[width=\linewidth]{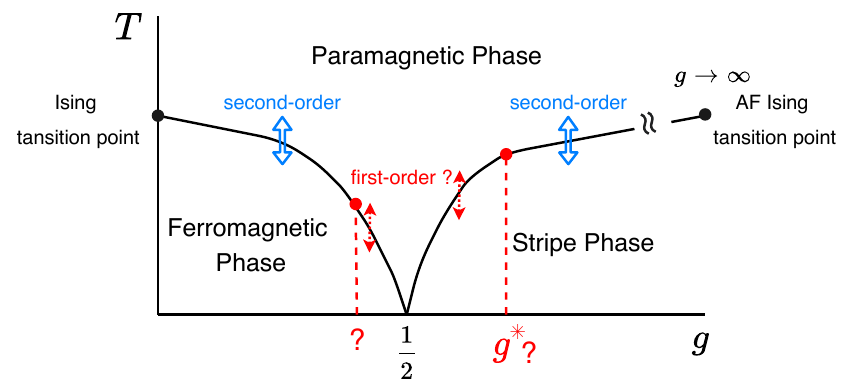}
 \caption{(Color online) A schematic phase diagram in the plane of temperature $T$ and coupling $g$ of $J_1$-$J_2$ Ising model in two dimensions. }
 \label{fig:phase_diagram}
\end{figure}

The order structure of this model is a ferromagnetic state with uniform magnetization (similar to the conventional Ising model) for a sufficiently small $g$, and a stripe state for a sufficiently large $g$.
The energy per spin of the ferromagnetic state in the ground state, $E_{\text{ferro}}$, and that of the stripe state, $E_{\text{stripe}}$, are given by
\begin{equation}
  \begin{cases}
    E_{\text{ferro}}  = -2 +2g,\\
    E_{\text{stripe}} = 2g
  \end{cases}
  \label{eqn:gs_energy}
\end{equation}
,respectively. These formulas imply that the ferromagnetic and stripe states are stable for $g<0.5$ and $g>0.5$, respectively, at least at zero temperature.  There is no issue regarding the order structure of low-temperature phases. However, the order of phase transitions and their universality classes have been discussed in the literature\cite{Kalz2012,Li2021}.
Fig.~{\ref{fig:phase_diagram}} displays the phase diagram of this model as indicated by the previous studies.

When the positive parameter $g$ varies, the characteristic of the phase transition at both the endpoints is known exactly. That is, $g=0$ is the conventional Ising model, and $g\to \infty$ corresponds to two mutually independent antiferromagnetic Ising models, both of which also belong to the universality class of the two-dimensional Ising model.
For $0<g<\infty$, exact analytical calculations are difficult, and the arguments have been based on mean-field calculations and numerical calculations. For example, the cluster mean-field analysis indicates the existence of a region of the first-order transition around $g=1/2$ for both $g<1/2$ and $g>1/2$\cite{Jin2013}.

Numerical studies indicate the following.
First, as mentioned above, the ferromagnetic phase transition occurs for $g<1/2$. Although many previous studies indicated that this phase transition belongs to the Ising universality class, the cluster mean-field analysis\cite{Jin2013} indicates the existence of the first-order transition around $g=1/2$. The transfer matrix calculation\cite{Hu2021} also shows a signature of a first-order transition in the region of $g\geq0.48$. That is, the issue of the order of phase transition for $g<1/2$ is not addressed completely.
Just at $g=1/2$, finite-size scaling of the peak temperature of the specific heat by MCMC indicates a phase transition at zero temperature\cite{Kalz2008}. Many studies\cite{Jin2013,Kalz2012,Hu2021} have concluded that no phase transition would occur at a finite temperature.

However, the characteristic of phase transition for $g>1/2$ is contentious.
The previous MCMC studies\cite{Jin2012,Jin2013,Kalz2012} contended that a certain critical value $g^*$ evaluated as $g^*=0.67(1)$ exists. It is a boundary of a first-order transition for $1/2<g<g^*$ and second-order transition for $g>g^*$.
They also indicated that this model belongs to the universality class of the Ashkin--Teller(AT) model \cite{Wiseman1993} for $g\geq g^*$.

One of the characteristics of the AT universality class is its weak universality\cite{MSuzuki1974}, where the critical exponents such as $\nu$ vary depending on the coupling constant $g$. However, the ratio of these exponents is constant as
\begin{equation}
  \left\{
  \begin{alignedat}{2}
    \frac{2-\alpha}{\nu}  &~=~& 2,\\
    \frac{\beta}{\nu}  ~~~&~= ~&\frac{1}{8},\\
    \frac{\gamma}{\nu}  ~~~&~=~& \frac{7}{4}.
  \end{alignedat}
  \right.
  \label{eqn:weak-universality}\end{equation}
In the AT universality class, the critical exponent $\nu$ varies with a lower bound of $\nu=2/3$, which corresponds to the four-state Potts universality class. In the $J_1$-$J_2$ Ising model, $\nu=1$ for $g \to \infty$. Therefore, the AT universality scenario asserts that $\nu$ varies in the range of  $2/3<\nu<1$ for $g^*<g<\infty$.

In contrast, other studies raise the question of the existence of a region of the first-order transition. For example, the previous study using iTEBD method\cite{Li2021} observed a second-order transition even at $g>0.54$ because there is no jump in the internal energy and other quantities at the transition temperature. The study also obtained a value of the central charge at $g=0.54$ close to that of the universality class of tricritical Ising model. This implies that the region of the first-order transition, if any, is narrower than expected from the MCMC results. This also indicates that it may be a second-order transition in all the regions with $g>1/2$. The iTEBD method calculates the thermodynamic limit under the approximation. Consequently, it is difficult to follow the influence of the approximation. Therefore, we study the phase transitions and critical phenomena of this model by large-scale HOTRG calculations. In particular, we also observe the finite-size behavior in the renormalization process to analyze it by finite-size scaling.

\section{Methods}
\label{sec:methods}
In this section, we describe the tenser network method (including its construction method), HOTRG method as an approximate contraction method, and impurity tensor method as a method for calculating certain physical quantities. The method for analyzing the physical observables obtained (finite-size scaling analysis (FSS)) is also described here.

\subsection{Tensor network}
\begin{figure}[ht]
 \centering
 \includegraphics[width=\linewidth]{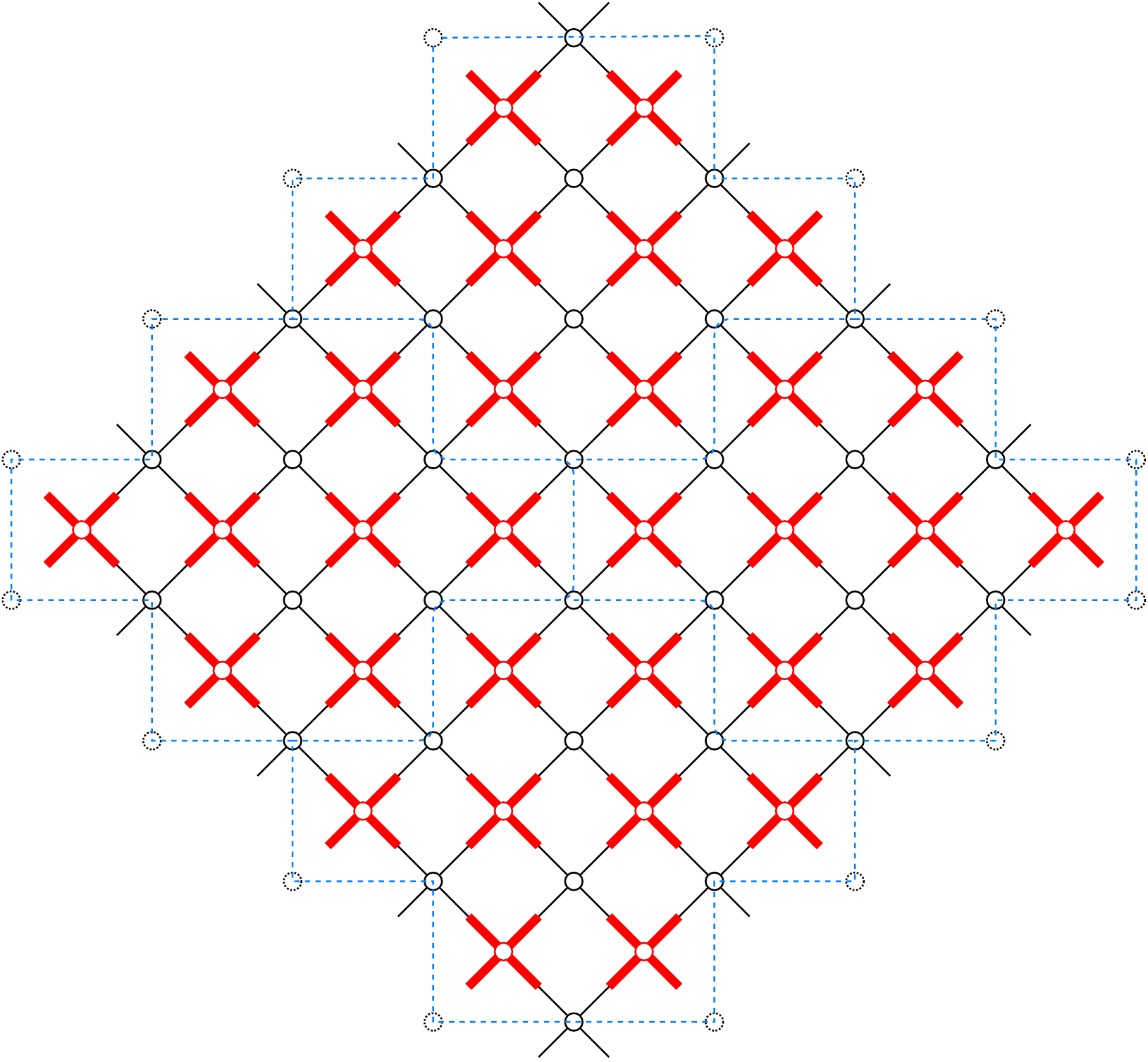}
 \caption{(Color online) An $8\times 8$ TN system with periodically aligned $4\times4$ TN systems surrounded by dotted lines. The thin (black) and thick (red) crosses represent tensors corresponding to $I$ and $T^1$, respectively. The circles on the $I$ tensors represent spins. The spins on the dotted line represent boundary spins, which impose periodic boundary conditions on the system in the sense that these are shared by several TNs.}
 \label{fig:TN4x4}
\end{figure}

There are several feasible settings for the TN representing the partition function of the $J_1$-$J_2$ Ising model. For example, the previous study \cite{Li2021} used a TN (hereafter referred to as type-I TN) with alternating tensors $I$ and $T^1$ defined as
\begin{align}
    I_{\sigma_a\sigma_b\sigma_c\sigma_d} = &\delta_{\sigma_a,\sigma_b}\delta_{\sigma_b,\sigma_c}\delta_{\sigma_c,\sigma_d},
    \label{eqn:TN1_a}\\
    T^1_{\sigma_a\sigma_b\sigma_c\sigma_d} = &e^{-\beta E_{\sigma_a\sigma_b\sigma_c\sigma_d}},
    \label{eqn:TN1_b}
\end{align}
with
\begin{align}
    E_{\sigma_a\sigma_b\sigma_c\sigma_d} = &-(\sigma_a\sigma_b +\sigma_b\sigma_c +\sigma_c\sigma_d +\sigma_d\sigma_a)/2\nonumber\\
    &+g(\sigma_a\sigma_c +\sigma_b\sigma_d),
    \label{eqn:local_energy}
\end{align}
where $\delta_{ij}$ is the Kronecker delta.
This corresponds to the partition function of a system for diagonally cutting a  square lattice as shown in Fig.~\ref{fig:TN4x4}. In this case, the dimension of each index of the initial tensor is $2$.

Another representation is to consider a TN (which is referred to as type-II TN) with a tensor $T^2$ defined by
\begin{align}
        T^2_{\bm{\sigma_{ab}}\bm{\sigma_{bc}}\bm{\sigma_{cd}}\bm{\sigma_{da}}} = &e^{-\beta E_{\sigma_a\sigma_b\sigma_c\sigma_d}},
  \label{eqn:TN2}
\end{align}
where $\bm{\sigma}_{ij} = (\sigma_i,\sigma_j)$.
This type-II TN is constructed on the face-centered lattice of a square lattice, which can also be used to represent the partition function of the system on the square lattice. The dimension of each index of the initial tensor is $4$.
Fig.~\ref{fig:TNa,b} shows diagrammatic representations of the two TNs mentioned above.
The filled circles and dashed lines in the diagram represent the spins and lattice of the original spin system.

\begin{figure}[ht]
 \centering
 \includegraphics[width=\linewidth]{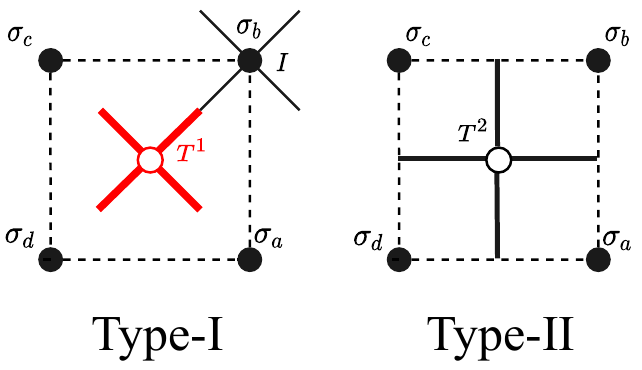}
 \caption{(Color online)Diagrammatic representations of TNs, (left) type-I for $I$ and $T^1$ given by Eq.~(\ref{eqn:TN1_a}) and (\ref{eqn:TN1_b}),  and (right) type-II for $T^2$ given by Eq.~(\ref{eqn:TN2})}
 \label{fig:TNa,b}
\end{figure}

The contraction of these TNs over the entire square lattice provides the partition function of the system. As explained in detail in the Appendix~\ref{sec:TN_selection}, the accuracy of calculations of physical quantities depends significantly on the contraction of these TNs. In the following, we mainly discuss the results obtained using the type-I TN defined by Eq.~(\ref{eqn:TN1_a}) and (\ref{eqn:TN1_b}), because this TN is more accurate. The system size $L$ is the length of one side of the square lattice on which TNs are defined. In the case of type-I TN, the number of spins of the original spin system is $N=L^2/2$.

\subsection{HOTRG}
Here, we describe HOTRG\cite{Xie2012} (the TN contraction method used in this study).
In HOTRG, two adjacent tensors are combined to form a tensor with an increased dimension. This is then renormalized by dimension reduction using singular value decomposition (SVD).

\begin{figure}[ht]
 \centering
 \includegraphics[width=\linewidth]{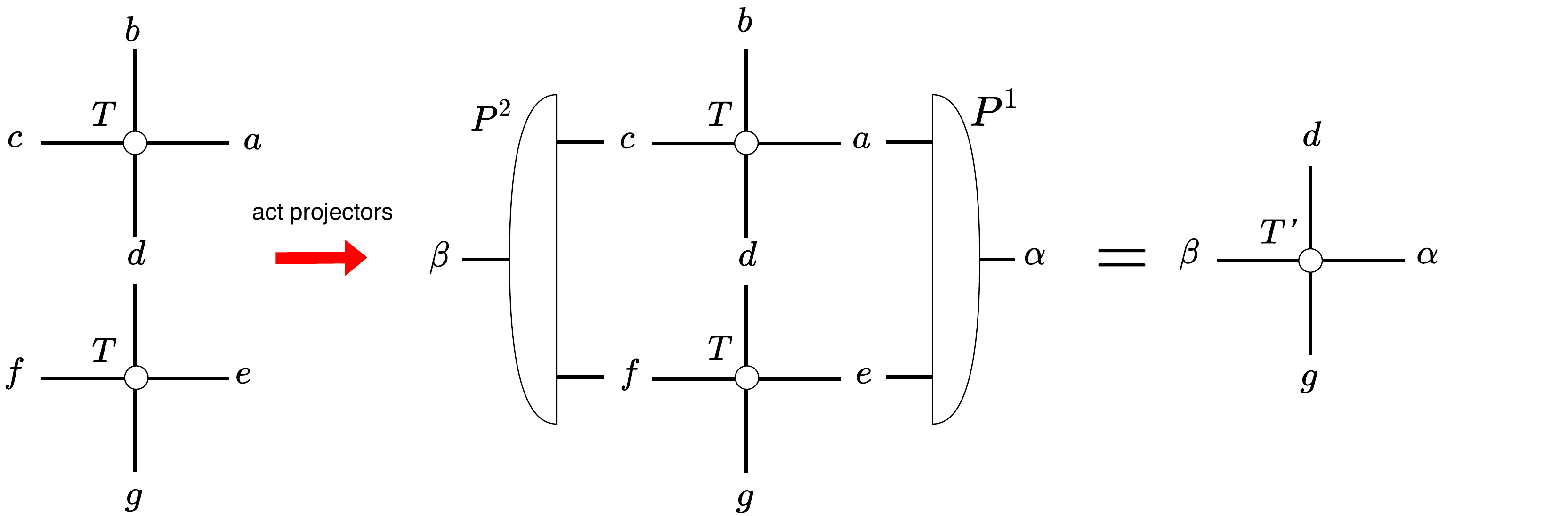}
 \caption{(Color online)Diagrammatic representation of a step of HOTRG renormalization}
 \label{fig:HOTRG1}
\end{figure}

First, for two tensors $T$ aligned along the $y$-axis (shown in the left view of Fig.~\ref{fig:HOTRG1})  a contraction of one of the indices yields a fourth-order tensor $\mathcal{T}$ given by
\begin{equation}
 \mathcal{T}_{ae,b,cf,g} = \sum_d^R T_{abcd}T_{edfg},
\label{eqn:connect}
\end{equation}
where $R$ is the dimension of an index of the tensor $T$. This operation increases the dimension of the $x$-axis legs $ae$ and $cf$ of $\mathcal{T}$ to $R^2$. The upper bound of the bond dimension is maintained constant at $D$ to reduce the computational complexity. This is achieved by acting on the projectors $P^1$ and $P^2$ as in
\begin{equation}
 T'_{\alpha,b,\beta,g} = \sum_{a,e,c,f} \mathcal{T}_{ae,b,cf,g}P^1_{ae,\alpha}P^2_{\beta,cf},
\label{eqn:renormalize}
\end{equation}
Eq.~(\ref{eqn:connect}) and Eq.~(\ref{eqn:renormalize}) together provide the transformation from $T$ to $T'$, or the renormalization transformation and are represented graphically as shown in Fig.~\ref{fig:HOTRG1}.
Here the projectors $P^1$ and $P^2$ are determined under the condition that $\mathrm{rank} (P^1P^2) \leq D$ by
\begin{equation}
  P^1,P^2 = \displaystyle\mathop{\arg \max}_{P^1,P^2} ||M^lM^r-M^lP^1P^2M^r||^2,
 \label{eqn:projector}
\end{equation}
where
 \begin{align}
  M^l_{bcfg,ae} &= \mathcal{T}_{ae,b,cf,g},\\
  M^r_{cf,gaeb} &= \mathcal{T}_{ae,b,cf,g}.
 \end{align}
Eq.~(\ref{eqn:projector}) is equivalent to a low-rank approximation of the matrix $M$ and is given by an SVD of $M^l,M^r$. See, for example, Ref.~\cite{Iino2019,Yoshiyama2020} for the derivation and other details.

\begin{figure}[ht]
 \centering
 \includegraphics[width=\linewidth]{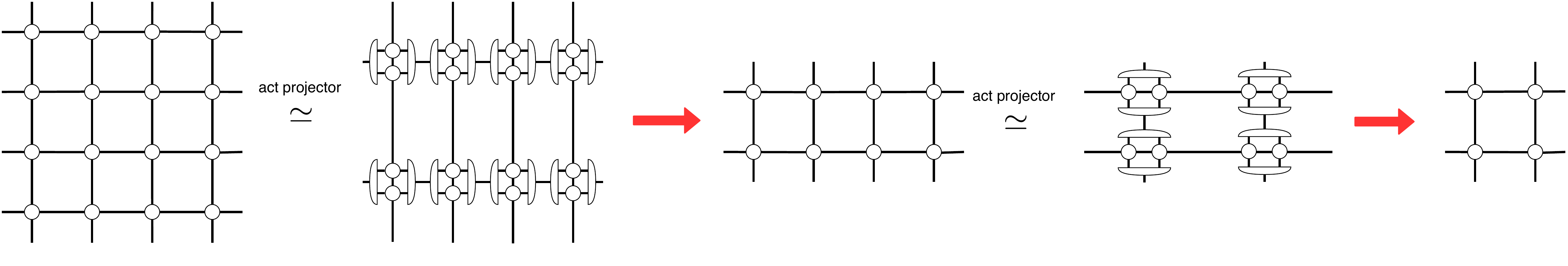}
 \caption{(Color online)Diagrammatic representation of tensor network transitions by HOTRG renormalization using projectors}
 \label{fig:HOTRG2}
\end{figure}

Similar to the renormalization in the $y$-direction, the renormalization in the $x$-direction is defined, and the renormalization procedures in two directions are performed alternately (see Fig.~\ref{fig:HOTRG2}).
The partition function $Z$ of the system of linear size $L=2^n$ under periodic boundary conditions is expressed as
\begin{equation}
    Z = \textrm{tTr } T^{(n)} \equiv \sum_{i,j} T_{ijij}^{(n)},
    \label{eqn:trace}
\end{equation}
where $\textrm{tTr } T^{(n)}$ is the trace of the tensor $T^{(n)}$ obtained by performing the renormalization $n$ times alternately in the $x$ and $y$-directions.

\subsection{impurity tensor}
The partition function $Z$ is calculated by the HOTRG method described above. In this section, we explain the impurity tensor method\cite{Gu2008,Morita2019}. It is a method for calculating moments of the order parameter and the internal energy that are derivatives of the free energy. The uniform magnetization $M$ and stripe magnetization $m$ (the order parameters for $g<1/2$ and $g>1/2$, respectively) are defined as
\begin{eqnarray}
    \begin{cases}
        M &\equiv ~\frac{1}{N}\displaystyle\sum_{x,y} \sigma_{x,y},\\
        m_v &\equiv ~\frac{1}{N}\displaystyle\sum_{x,y} (-1)^x\sigma_{x,y},\\
        m_h &\equiv ~\frac{1}{N}\displaystyle\sum_{x,y} (-1)^y\sigma_{x,y},\\
        m &\equiv ~m_v +m_h,
    \end{cases}
\end{eqnarray}
,respectively. Here, $\sigma_{x,y}$ represents a spin at $(x,y)$ (the coordinates in a square lattice).

In this method, a tensor specific to the physical quantities to be calculated is defined by multiplying each component of the local tensor by its local physical quantity. Such a tensor is called an impurity tensor.
We first introduce local stripe magnetizations $m_h$ and $m_v$ along the $x$ and $y$ directions, respectively, as
\begin{align}
    m_h(x,y) &= \frac{(-1)^y}{4}(\sigma_{x,y}-\sigma_{x,y+1}
    -\sigma_{x-1,y+1}+\sigma_{x-1,y}),\\
    m_v(x,y) &= \frac{(-1)^x}{4}(\sigma_{x,y}+\sigma_{x,y+1}
    -\sigma_{x-1,y+1}-\sigma_{x-1,y}).
\end{align}
where the four spins $\sigma_{x,y},\sigma_{x,y+1},\sigma_{x-1,y+1}$ and $\sigma_{x-1,y}$ in the tensor correspond to the four spins $\sigma_a$, $\sigma_b$, $\sigma_c$, and $\sigma_d$, respectively, in Fig.~\ref{fig:TNa,b}.
In addition to the local energy in Eq.~(\ref{eqn:local_energy}), the local order parameter for the stripe phase is defined by
\begin{equation}
     m(x,y) =  m_h(x,y) +m_v(x,y).
\end{equation}
The averages of these local physical quantities $m(x,y)$, $m_h(x,y)$, and $m_v(x,y)$ yield the macroscopic physical quantities $m$, $m_h$ and $m_v$.
The impurity tensors of the local physical quantities $m$ and $E$, corresponding to the tensor in Eq.~(\ref{eqn:TN2}), are given by
\begin{align}
     (S_k^m(x,y))_{abcd} &= m(x,y)^ke^{-\beta E_{abcd}},\label{eqn:impurity_m}\\
     (S_k^E)_{abcd} &= E_{abcd}^k e^{-\beta E_{abcd}}.
      \label{eqn:impurity_E}
\end{align}
, respectively.
The procedure defined by Eq.~(\ref{eqn:connect}) and (\ref{eqn:renormalize}) renormalizes two tensors $T$ aligned in the $y$-direction into one tensor $T'$. It is formally denoted by
\begin{equation}
  T'\leftarrow  TT.
  \label{eqn:recursion_T}
\end{equation}
The impurity tensors defined in Eq.~(\ref{eqn:impurity_m}) and (\ref{eqn:impurity_E}) are renormalized similarly as in Eq.~(\ref{eqn:connect}) and (\ref{eqn:renormalize}). The renormalization procedure is expressed in a recursive manner using the formal expressions as in Eq.~(\ref{eqn:recursion_T}), as follows:
\begin{equation}
  \begin{split}
  S'_1  \leftarrow &\frac{1}{2^1}(ST +TS),\\
  S'_2 \leftarrow &\frac{1}{2^2}(S_2T +2SS +TS_2),\\
  S'_3 \leftarrow &\frac{1}{2^3}(S_3T +3S_2S +3SS_2+TS_3),\\
  &\vdots\\
  S'_k \leftarrow &\frac{1}{2^k}\sum_{i=0}^k \binom{k}{i} S_{k-i} S_i,
  \label{eqn:recursion_S}
  \end{split}
\end{equation}
where $S_0\equiv T, ~S_1\equiv S$ and $\binom{k}{i}$ is a binomial coefficient.

Using the renormalized impurity tensor $S_k$ calculated thus, the higher-order moments of the physical quantity per spin for a system of size $L=2^n$ are evaluated by
\begin{equation}
    \langle E^k\rangle = \frac{\textrm{tTr }  {S_k^{E}}^{(n)}}{\textrm{tTr } T^{(n)}}.
    \label{eqn:moment}
\end{equation}
where ${S_k^{E}}^{(n)}$ is formed by renormalizing ${S_k^{E}}$ for $n$ times in the $x$- and $y$-directions according to Eq.~(\ref{eqn:recursion_S}).
The $k$-th power of the order parameter is also expressed with the impurity tensor as
\begin{equation}
    \langle m^k\rangle = \frac{\textrm{tTr }  {S_k^{m}}^{(n)}}{\textrm{tTr } T^{(n)}}.
    \label{eqn:moment_m}
\end{equation}
The specific heat $C$ and Binder parameter $R_4$ are defined from these higher order moments by
\begin{equation}
  \begin{cases}
    C = N(\langle E^2\rangle -\langle E\rangle^2),\\
    R_4 = \displaystyle\frac{\langle m^4\rangle}{\langle m^2\rangle^2}.
  \end{cases}
  \label{eqn:C-R_4}
\end{equation}
, respectively. When the transition is of second order, the specific heat diverges algebraically, and $R_4$ increases from $1$ to $3$ at the transition temperature $T_c$. In contrast, for a first-order transition, both $C$ and $R_4$ are expected to diverge of the $\delta$-function type.

\subsection{Finite size scaling}
We employ finite-size scaling for the results obtained by HOTRG to study critical phenomena.
Assuming a second-order transition, a finite size scaling (FSS) form of a critical physical quantity $X$ is given by\cite{Binder1981}
\begin{equation}
     X(T,L) = L^{\phi_X}f_{X}((T-T_c)L^{1/\nu}),
     \label{eqn:FSS_Binder}
\end{equation}
where $\phi_X$ is a scaling dimension and $f_{X}$ is the universal scaling function for $X$. Because the Binder parameter $R_4$ is a dimensionless quantity,  its scaling dimension $\phi_{R_4}=0$, and its FSS is effective for evaluating the exponent $\nu$.
The temperature derivative of $R_4$ at $T_c$ (evaluated by numerical differentiation in this study)  is also effective. Its scaling form is given by
\begin{equation}
    \left.\frac{dR_4}{dT}\right|_{T=T_c} \propto L^{1/\nu}.
  \label{eqn:dRdT_FSS}
\end{equation}
The scaling of this quantity for the first-order transition is expected to be $\nu=1/d$\cite{Fisher1982}. Here, $d$ is the spatial dimension.

Similarly, the scaling dimension of the squared order parameters $\langle M^2\rangle$ and  $\langle m^2\rangle$ is $\phi_{M^2}=-2\beta/\nu$. Moreover, its scaling form at $T_c$ is given by
 \begin{equation}
     \left.\langle m^2 \rangle \right|_{T=T_c} \propto L^{-2\beta/\nu},
     \label{eqn:FSS_m2}
\end{equation}
where $2\beta/\nu=d-2+\eta$ from the scaling relation.
In the thermodynamic limit, the inverse of the logarithmic derivative of $\langle M^2 \rangle $ conforms to
\begin{equation}
  \left(\frac{\partial}{\partial T}\log \langle M^2 \rangle \right)^{-1}
  =\begin{cases}
     \frac{T-T_c}{2\beta} &(T<T_c),\\
     -\frac{T-T_c}{\gamma} &(T>T_c).
   \end{cases}
   \label{eqn:logm_scaling}
\end{equation}
This relationship holds for $\langle m^2 \rangle$. The slope of the temperature dependence of this quantity above and below the transition temperature represents the critical exponents and is effective for their estimation.

Furthermore, the scaling dimension of the specific heat is given by  $\phi_C=\alpha/\nu$. In the analysis of the specific heat of finite-size systems, the divergent exponent of the peak value of the specific heat is generally evaluated as
\begin{equation}
    C_{\rm max}(L)\simeq L^{\alpha/\nu},
    \label{eqn:ctemp_peak}
\end{equation}
where $C_{\rm max}$ is the peak value of the specific heat of size $L$ as a function of temperature. The correlation length exponent $\nu$ is evaluated from the scaling form of the peak temperature $T_{\rm max}$ with $L$ given by
\begin{equation}
  T_{\rm max} (L)-T_{\rm max}(\infty) \propto L^{-1/\nu},
  \label{eqn:ctemp_FSS}
\end{equation}
where $T_{\rm max}(\infty)$ is the transition temperature defined in the thermodynamic limit.

\section{Numerical results}
\label{sec:results}
In this section, we present the numerical results obtained by our HOTRG calculations for the $J_1$-$J_2$ Ising model with the parameter $g$ based on the type-I TN explained in the previous section.
\subsection{Ising universality class for $g<1/2$}
\label{sec:result_g<1/2}
\begin{figure}
 \centering
 \includegraphics[width=\linewidth]{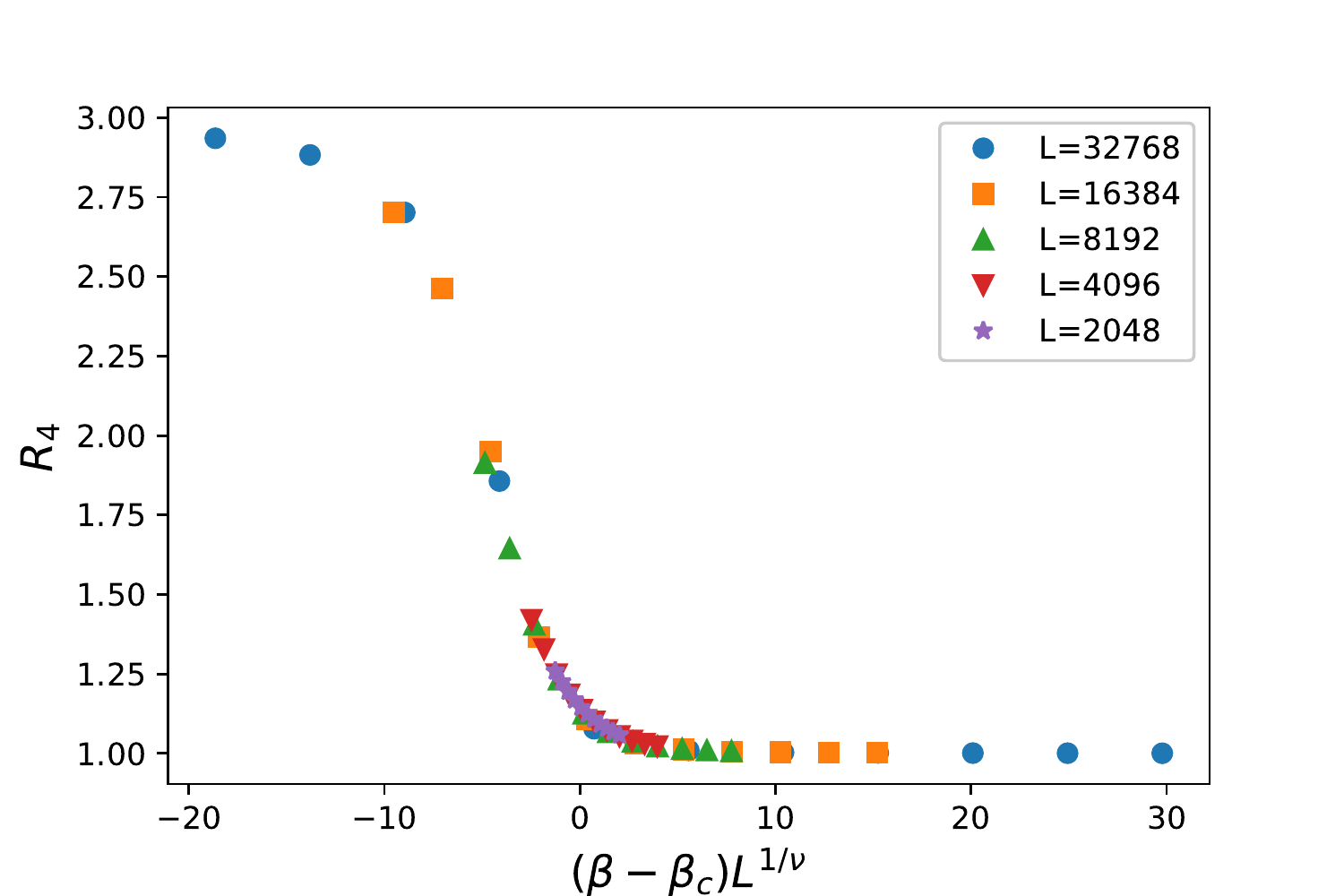}
 \caption{(Color online)Finite-size scaling plot of the Binder parameter $R_4$ at $g=0.49$ with $D=32$. This scaling plot is obtained by $\beta_c=2.65277(2)$ and $\nu=1.03(5)$. }
 \label{fig:g049_R4_FSS}
\end{figure}

First, we show the HOTRG results for $g<1/2$, where a ferromagnetic phase with uniform magnetization is expected to occur.
The FSS plot of the Binder parameter at $g=0.49$ obtained by the Bayesian scaling analysis\cite{Harada2015} is shown in Fig.~\ref{fig:g049_R4_FSS}. It yields $\beta_c=2.65227(2)$ and $\nu=1.03(5)$. This is consistent with the Ising universality class.

\begin{figure}
 \centering
 \includegraphics[width=\linewidth]{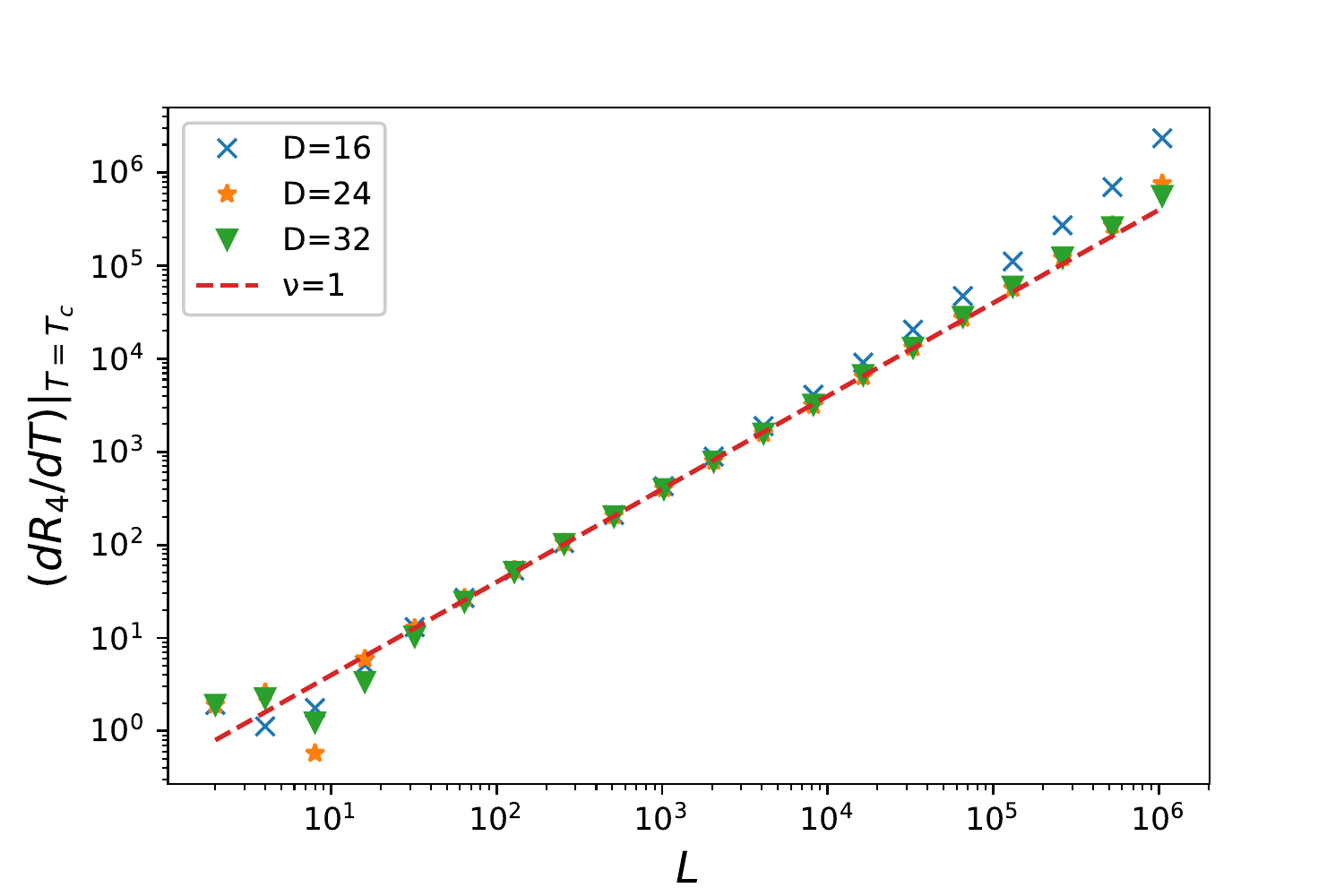}
 \caption{(Color online)Size dependence of the temperature derivative of the Binder parameter, $\left.\frac{dR_4}{dT}\right|_{T=T_c}$ at $T_c$ and $g=0.49$ with certain values of $D$. The dashed line represents a straight line with the slope expected from the Ising universality.  }
 \label{fig:g049_dR4dT_FSS}
\end{figure}

Fig.~\ref{fig:g049_dR4dT_FSS} presents the FSS of the temperature derivative of $R_4$ given by Eq.~(\ref{eqn:dRdT_FSS}). It verifies that the size region that follows a scaling with $\nu=1.0$ becomes wider with an increase in $D$.
In the context of TN, the value of $D$ determines an upper bound on the correlation length that can be simulated and is considered to correspond to a certain length scale\cite{Ueda2014}. The result observed here is in good agreement with this picture.
For $32<L<65536$, where $(dR_4/dT)|_{T=T_c}$ with $D=32$ appears to follow the power of $L$ well, the result of fitting to the power law yields $\nu=1.004(5)$. This again indicates that it belongs to the Ising universality.

\begin{figure}
 \centering
 \includegraphics[width=\linewidth]{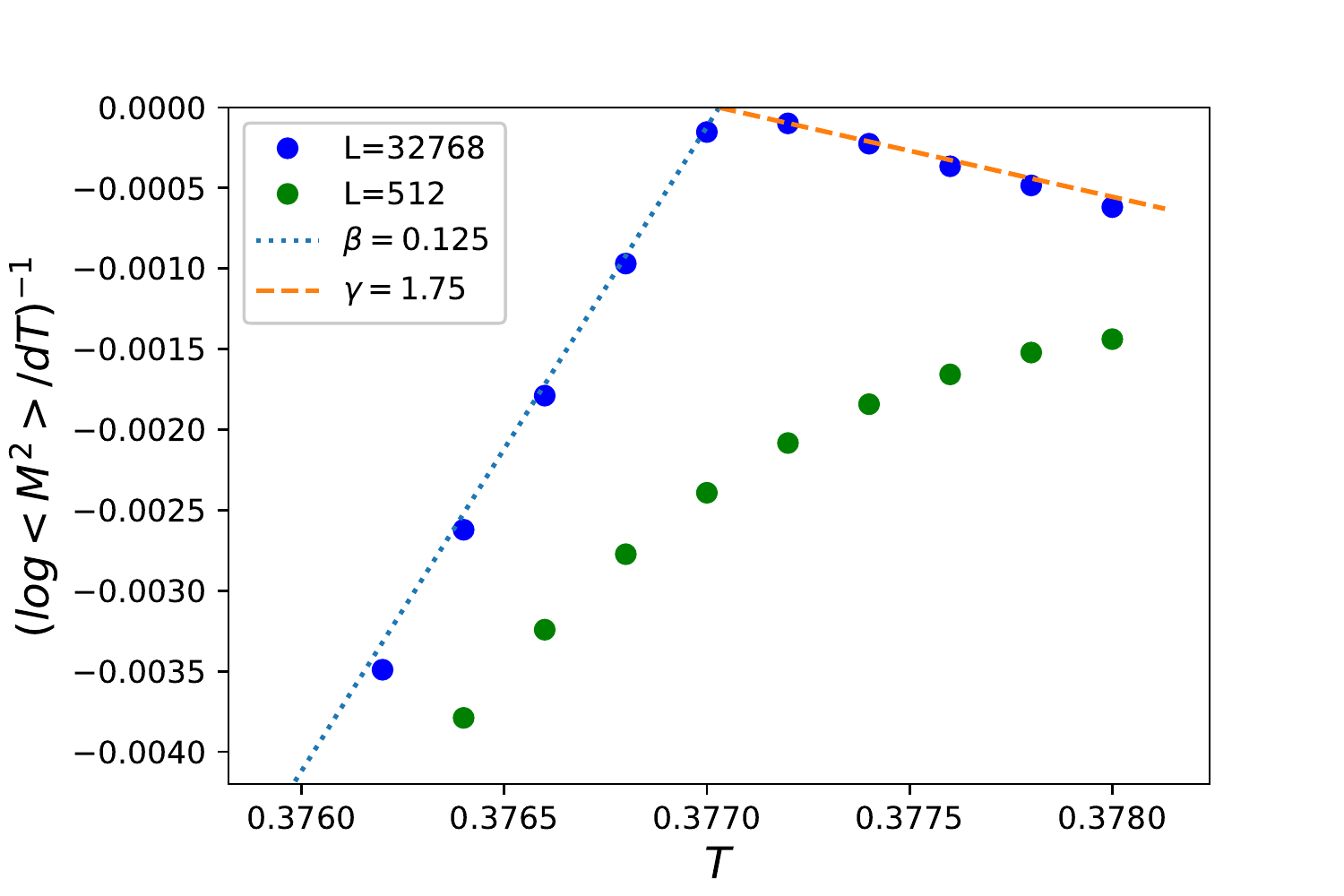}
 \caption{(Color online)
 Temperature dependence of the inverse of the logarithmic derivative of $\langle M^2 \rangle$ of the $J_1$-$J_2$ Ising model at $g=0.49$. The (blue) circle represents the result for $L=32768$ and the (green) square represents the result for $L=512$. The dotted and dashed lines represent scaling with $\beta=1/8$ and $\gamma=7/4$, respectively. The transition temperature is estimated to be $T_c=0.3770$.}
 \label{fig:g049_dlogM}
\end{figure}

We also verify that the critical exponents $\beta$ and $\gamma$ are consistent with the Ising universality using Eq.~(\ref{eqn:logm_scaling}). Fig.~\ref{fig:g049_dlogM} shows the temperature dependence of $\left(\frac{\partial}{\partial T}\log \langle M^2 \rangle \right)^{-1}$ for $L=512$ and $32768$ at $g=0.49$ and $D=32$. The two dotted lines above and below the transition temperature are straight lines representing the values $\beta=1/8$ and $\gamma=7/4$, respectively, of the critical exponents of the Ising universality. The result for $L=32768$ agrees well with the two dotted lines. This supports the assertion that the model at $g=0.49$ belongs to the Ising universality. This figure demonstrates that this scaling is not verified sufficiently for sizes of approximately $L=512$, which can be accessed by MCMC. This again demonstrates the advantage of the HOTRG method for large-size calculations.

Because the transition temperature is lower, it is difficult to calculate $g>0.49$ owing to the numerical accuracy.  However, in this study, the phase transition at $g=0.49$ is verified to belong to the Ising universality class. This indicates that most of the phase transitions in the region $g<1/2$ are covered by the same universality class.

\subsection{gauge invariant quantity}
\label{sec:X}

To determine the number of internal degrees of freedom of the tensor that is renormalized $n$ times, TRG studies generally measured the gauge invariant quantity $X$ defined by\cite{Gu2009}
\begin{equation}
    X^{(n)}\equiv \frac{\left(\displaystyle\sum_{r,u}T^{(n)}_{ruru}\right)^2}{\displaystyle\sum_{r,u,l,d}T^{(n)}_{rulu}T^{(n)}_{ldrd}}.
    \label{eqn:X}
\end{equation}
This quantity takes $1$ in a disordered phase such as the paramagnetic phase, and the value of the number of states in the ordered phase. It is used as a method to detect phase transitions with the aid of the almost discontinuous jump at the transition temperature for large system sizes \cite{Li2022}.

\begin{figure}[ht]
 \centering
 \includegraphics[width=\linewidth]{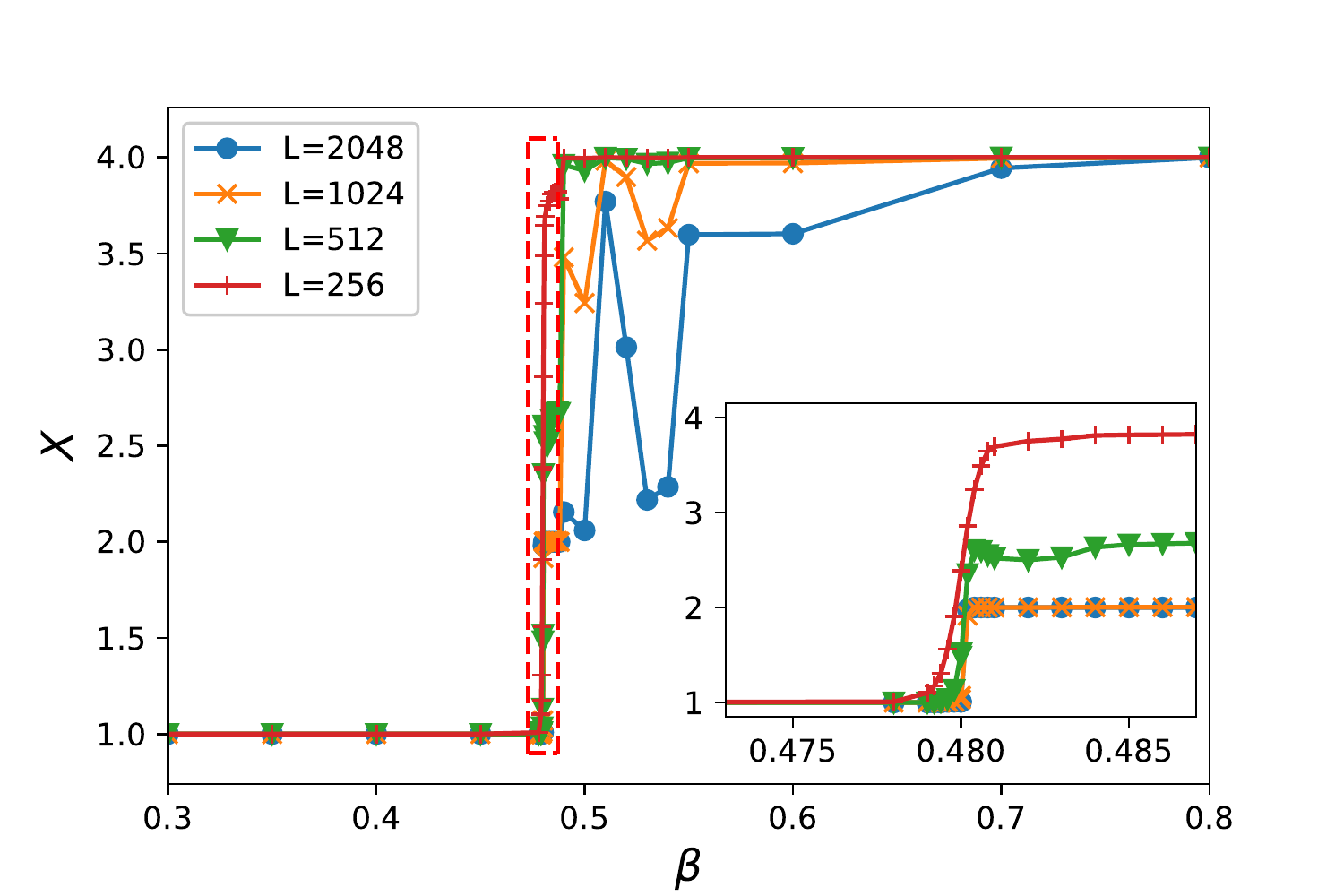}
 \caption{(Color online)Inverse-temperature dependence of $X$ for the $J_1$-$J_2$ Ising model at $g=1.0$ and with $D=32$. The inset presents an enlarged view of the region enclosed by the red dotted line in the main figure.
 }
 \label{fig:g1.0_X}
\end{figure}

In the $J_1$-$J_2$ model, the quantity $X$ is expected to vary from $1$ to $2$ for $g<1/2$ and from $1$ to $4$ for $g>1/2$ at the transition temperature with a decrease in temperature. Although such behavior is verified for $g<1/2$, not shown here, it does not follow the naive expectation for $g>1/2$. As shown in Fig.~\ref{fig:g1.0_X}, $X$ adopts the value $4$ for small sizes. However, for sizes $L$ larger than approximately $1000$, a plateau is observed at $X=2$ immediately below the transition temperature before the expected value $4$ is attained at a low temperature. This may be interpreted as another intermediate phase between the high-temperature paramagnetic phase and low-temperature stripe phase. However, the temperature at which $X$ varies from $2$ to $4$ depends significantly on the size $L$. This indicates that the plateau at $X=2$ is owing to a numerical artifact caused by the amplification of approximation errors in the renormalization process by iterations rather than a thermodynamic phase transition from the intermediate phase to the low-temperature phase. A similar artifact-like behavior is observed in certain physical quantities shown below. It is considered to be a result of the effect of a large number of renormalizations. It should be noted that to estimate the critical exponents, one should use the system sizes that do not generate such artifact effects.

\subsection{$1/2<g<g^*$:first order transition}
\label{sec:result_g<g^*}
\begin{figure}[ht]
 \centering
 \includegraphics[width=\linewidth]{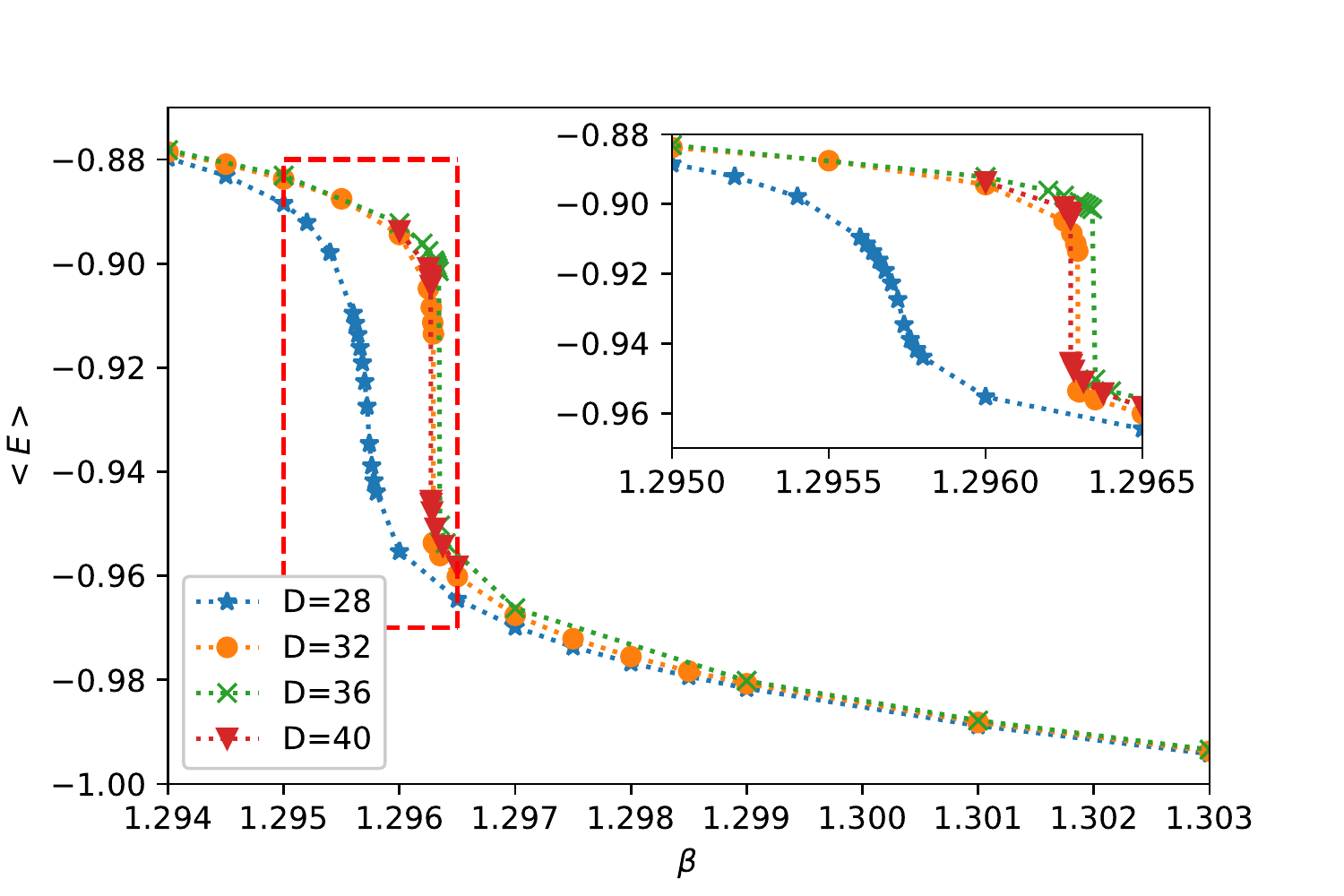}
 \caption{(Color online)
 Inverse-temperature dependence of the internal energy $\langle E\rangle$ of the system at $g=0.55$ with certain values of $D$ and $L=32768$. The inset shows an enlarged view near the transition temperature.
 }
 \label{fig:g=0.55_E_Ddep}
\end{figure}

Next, we discuss the system at $g=0.55$. A first-order transition is asserted to occur here in the previous MCMC studies\cite{Jin2012,Jin2013} and a second-order transition in the previous iTEBD study\cite{Li2021}.
Fig.~\ref{fig:g=0.55_E_Ddep} presents the internal energy $\langle E\rangle$ for different inverse temperatures obtained by our HOTRG calculations with certain values of $D$.
It is observed that $\langle E \rangle$ is nearly convergent at $D\geq 32$ except near the transition temperature. Near this temperature, the internal energy for $D=28$ is continuous as a function of the inverse temperature. This indicates a second-order transition. Meanwhile, the results for $D\geq 32$ display a sharp jump at the transition temperature at $\beta_c\simeq 1.2963$. This supports a first-order transition.

\begin{figure}[ht]
 \centering
 \includegraphics[width=\linewidth]{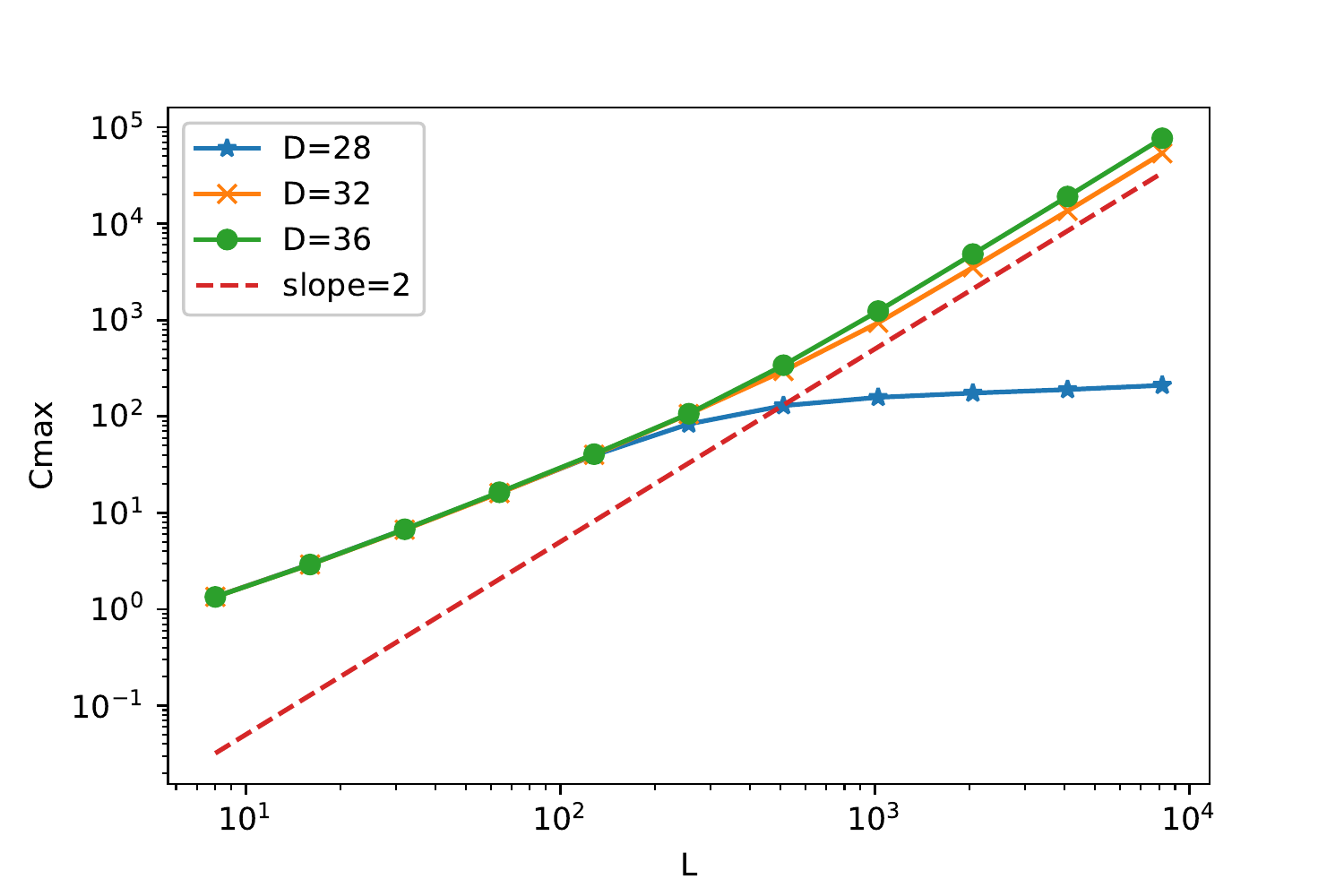}
 \caption{(Color online)
 System size $L$ dependence of the peak value of the specific heat $C_{\text{max}}$ at $g=0.55$ with $D=28$, $32$ and $36$. The dotted line represents a power law as $L^2$.
 }
 \label{fig:g=0.55_C}
\end{figure}

The size dependence of the peak height of the specific heat is shown in Fig.~\ref{fig:g=0.55_C}. For a relatively small $D$, i.e., $D=28$, the peak value of specific heat saturates at a certain size presumably caused by the finite $D$ effect. However, for a large $D$, it continues to grow following the $L^2$ scaling. This is characteristic of the first-order transition. Thus, our HOTRG results at $g=0.55$ for the energy and specific heat indicate a first-order phase transition for a large $D$. This is the more accurate calculation, although it appears to be a second-order phase transition for a smaller $D$.
This conclusion is in contrast to that in the previous study using the iTEBD method\cite{Li2021} at $g=0.55$. That study asserted that the energy varies continuously under the approximation of finite bond dimension. Our results indicate that the likely effects of the bond dimensions need to be carefully examined in the previous study as well.

\begin{figure}[ht]
 \centering
 \includegraphics[width=\linewidth]{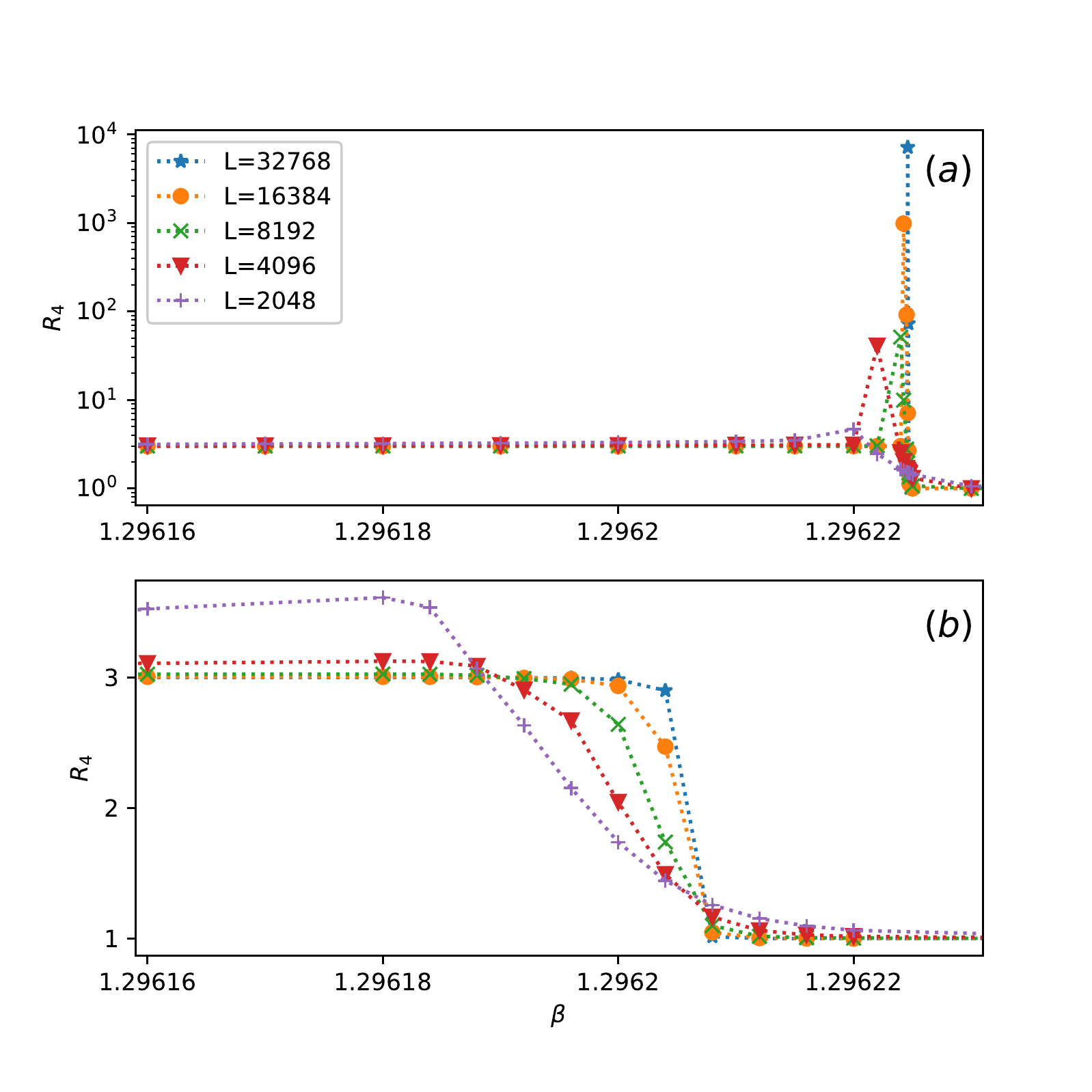}
 \caption{(Color online)
 Inverse temperature $\beta$ dependence of the Binder parameter $R_4$ at $g=0.55$ for (a)$D=40$ and (b)$D=36$}
 \label{fig:g=0.55_R4}
\end{figure}

However, this is not the case for the Binder parameter. Fig.~\ref{fig:g=0.55_R4}(a) shows the inverse-temperature dependence of the Binder parameter at $D=40$. It exhibits a sharp peak near the transition temperature, indicating the first-order transition. However, such a behavior disappears and appears with an increase in $D$ and therefore, is unstable with respect to $D$. For example, a different behavior is observed for $D=36$, as shown in Fig.~\ref{fig:g=0.55_R4}(b). The peak of the Binder parameter still remains for sizes that can be calculated with MCMC \cite{Jin2012}. However, such first-order transition-like behavior disappears as the system size increases. Eventually, the Binder parameter decreases monotonically with $\beta$.

\begin{figure}[ht]
 \centering
 \includegraphics[width=\linewidth]{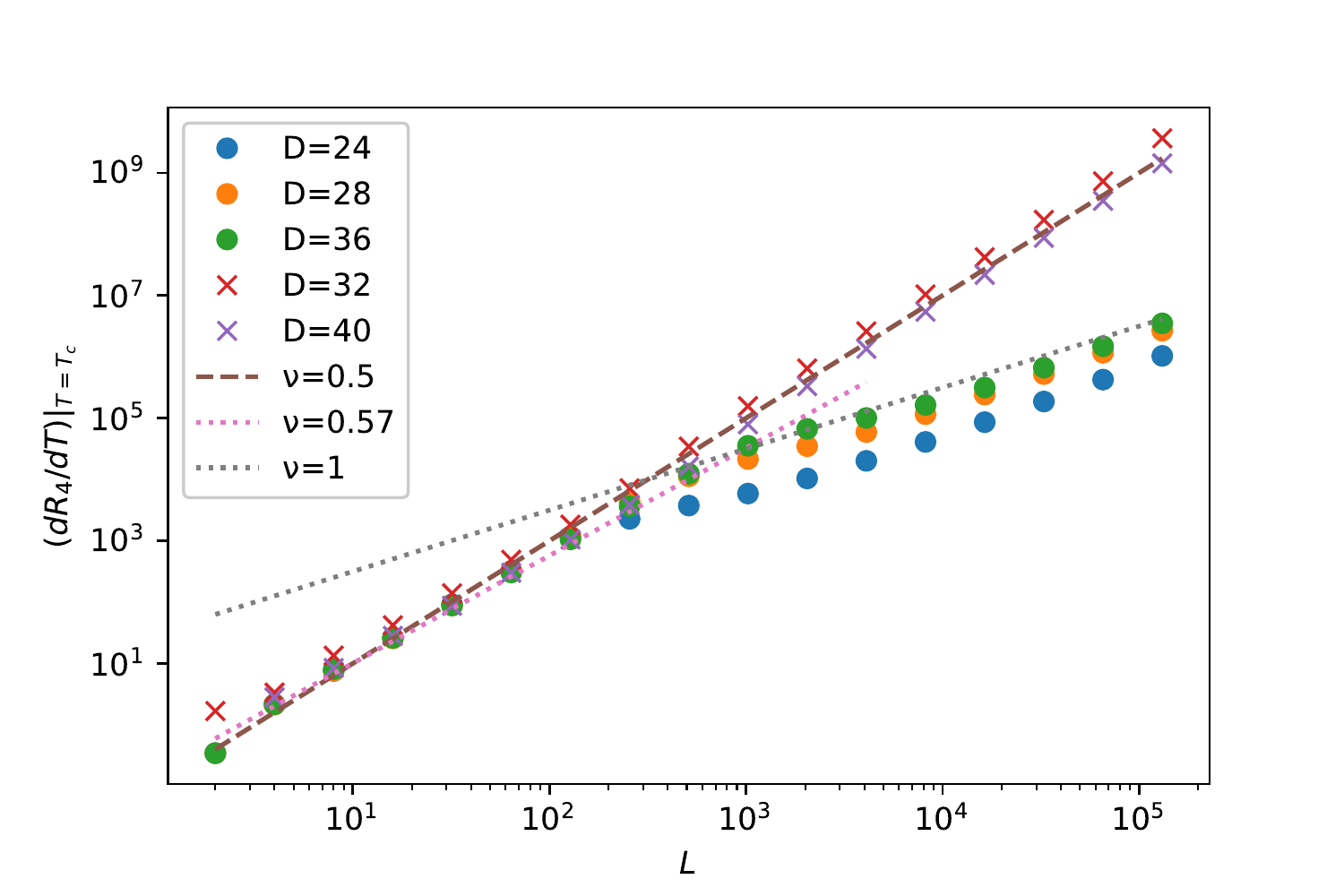}
 \caption{(Color online)System-size dependence of $(dR_4/dT)|_{T=T_c}$ at $g=0.55$ with certain values of $D$. The straight lines represent power laws with the exponent $\nu=1.0$ for Ising universality, $\nu=0.57$ for a non-trivial case, and $\nu=1/2$ for a first-order transition in two dimensions. }
 \label{fig:g=0.55_dR4dT_FSS}
\end{figure}

We also calculate the temperature derivative of the Binder parameter at the transition temperature for $g=0.55$, which is expected to follow the power law of $L$ as in Eq.~(\ref{eqn:dRdT_FSS}). Its exponent depends on the order of the transition. As shown in Fig.~\ref{fig:g=0.55_dR4dT_FSS}, it is observed that the $L$ dependence of $(dR_4/dT)|_{T=T_c}$ at $g=0.55$ also depends on $D$, corresponding to the $D$ dependence of the $R_4$.
For example, for $D=32$ and $40$ ,where $R_4$ shows the first-order transition-like behavior, $(dR_4/dT)|_{T=T_c}$ follows $\nu=1/d$ with $d$ being spatial dimensions up to sufficiently large sizes. This is consistent with the first-order transition. Meanwhile, for $D=24$, $28$, and $36$ where $R_4$ shows the second-order transition-like behavior, its derivative follows the power law with a nontrivial critical exponent, although the behavior deviates from the power law in the order of decreasing $D$. The critical exponent is obtained as $\nu=0.57(1)$ by linear regression from the data following the power law. The finite $D$ effect causes a deviation from this power law and follows $\nu\simeq 1$ for a sufficiently large $L$.
This behavior can be considered as an example of the Ising-like behavior observed after repeated renormalization as described in Sec.\ref{sec:X}

The above results verify that at $g=0.55$, certain physical quantities such as $\langle E\rangle$, $C$, and $R_4$ exhibit indications of a first-order transition. These quantities may also show a second-order transition-like behavior when $D$ is modified. For example, $\langle E\rangle$ and $C$ behave similar to a second-order transition only when $D$ is small, and to a first-order transition when $D$ is large. Although the behavior of $R_4$ varies rather sensitively to $D$, such behaviors are considered to be a result of only the approximation of a finite $D$. It is strongly indicated that this model for $g=0.55$ shows the first-order transition for $D\to\infty$.

\subsection{Edge of the first-order transition: critical-end point $g^*$}
\label{sec:g^*}
\begin{figure}[ht]
 \centering
 \includegraphics[width=\linewidth]{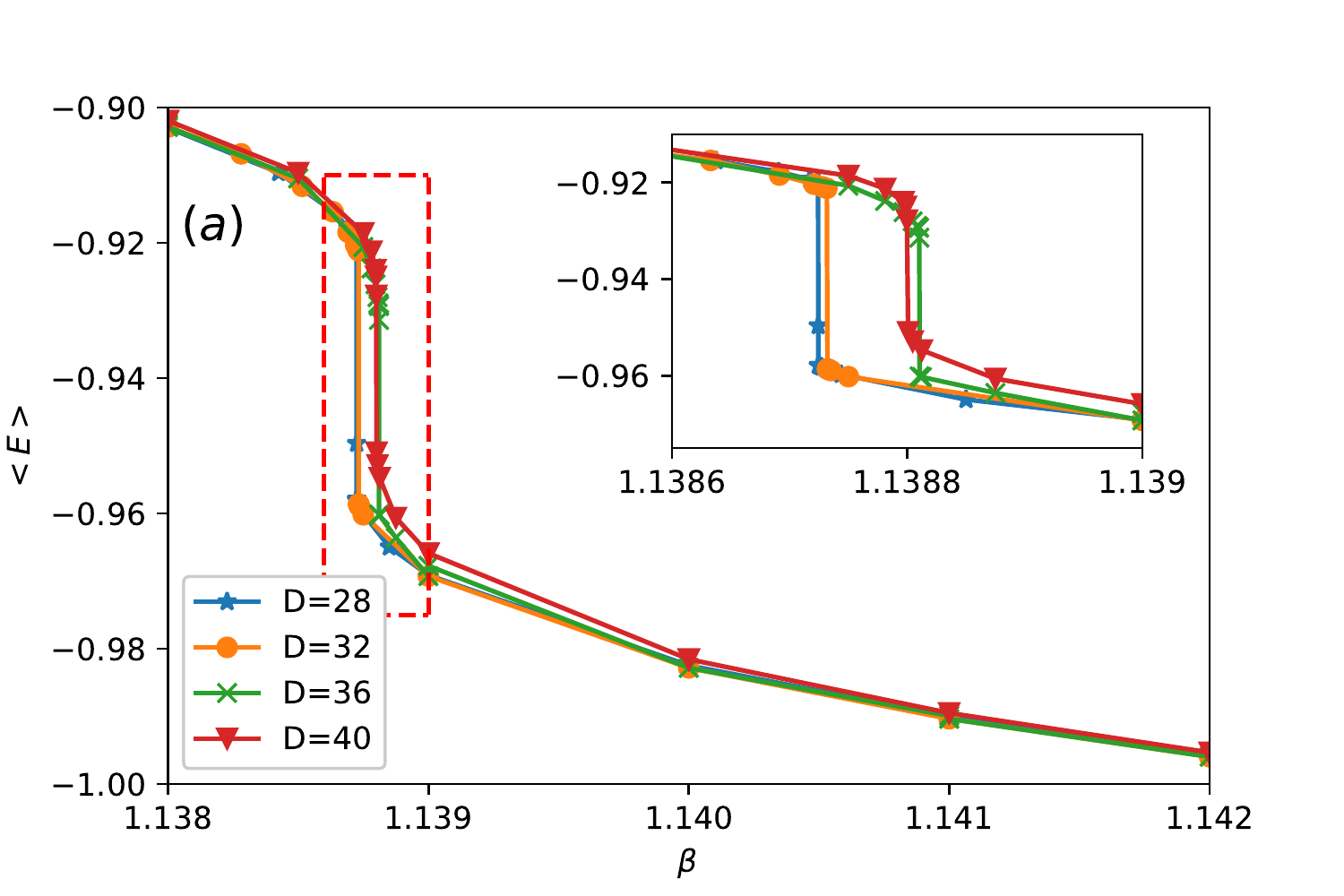}
 \centering
 \includegraphics[width=\linewidth]{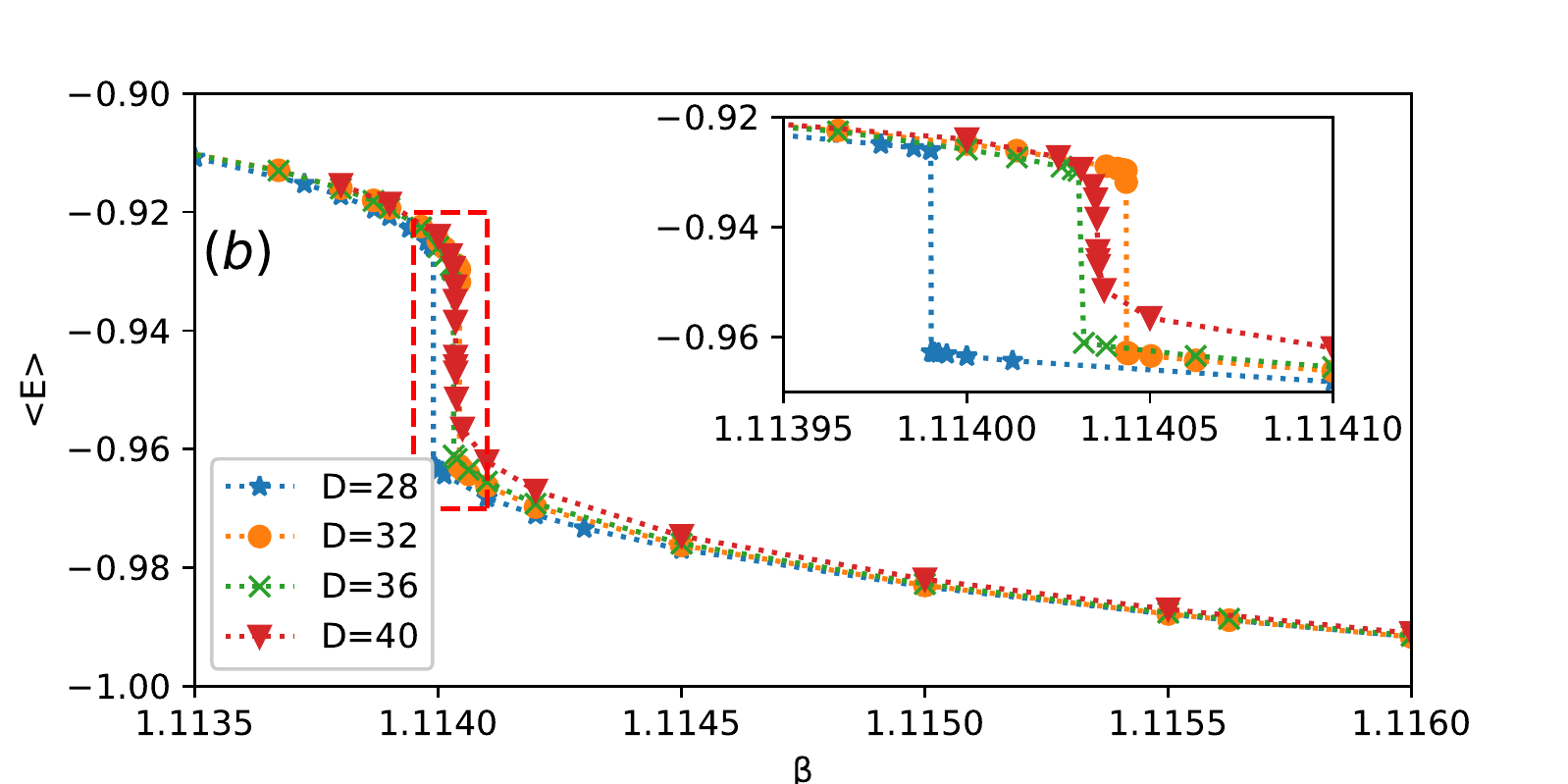}
  \caption{(Color online)Inverse-temperature dependence of the internal energy for certain values of $D$ at $g=0.575$ (a) and $g=0.58$ (b). The system size is $L=32768$.  The insets are enlarged views around the transition temperature.}
 \label{fig:g0.575_E}
\end{figure}

One of the issues to be resolved in this model is to determine the value of the boundary $g^*$ between the first-order and second-order transitions for a varying $g$. Here, we investigate in detail the energy jump at the transition temperature. Fig.~\ref{fig:g0.575_E} shows the inverse-temperature dependence of the energy at $g=0.575$ and $0.58$ for different values of $D$. For a small $D$, i.e., $D=28$, $32$, or $36$, the energy exhibits first-order transition-like behaviors for both $g=0.757$ and $g=0.58$. Meanwhile, for $D=40$, it remains discontinuous at $g=0.575$ and becomes continuous at $g=0.58$.
 Such a $D$-dependence, which transforms from discontinuous to continuous with an increase in $D$, is a behavior that is the converse of the transformations from continuous to discontinuous observed at $g=0.55$.
 Although we cannot completely exclude the possibility that a further increase in $D$ would again show a discontinuous jump at $g=0.58$, the result for $D=40$ indicates that $0.575<g^*<0.58$. This value of $g^*$ is considerably smaller than that estimated by the previous work with MCMC.

In general, the higher the order of the derivative of the free energy, the lower the accuracy of the approximation in the calculations of impurity tensors. Hence, $\langle E \rangle$ is considered to be more accurate than $\langle m^2 \rangle$. Therefore, based on the behavior of $\langle E \rangle$ at $D=40$, the order of the phase transition at $g=0.575$ is considered the first-order transition, and we conclude $g^*=0.58$ at this time.
However, the evaluated value $g^*$ of this boundary still depends on the value of $D$ that we can calculate. It also appears to behave differently depending on the physical observables.
Therefore, the value $g^*=0.58$ can still contain uncertainty that needs to be investigated further.

\subsection{$g\geq g^*$: Universality class}
\label{sec:result_g>g^*}

\begin{figure}[ht]
 \centering
 \includegraphics[width=\linewidth]{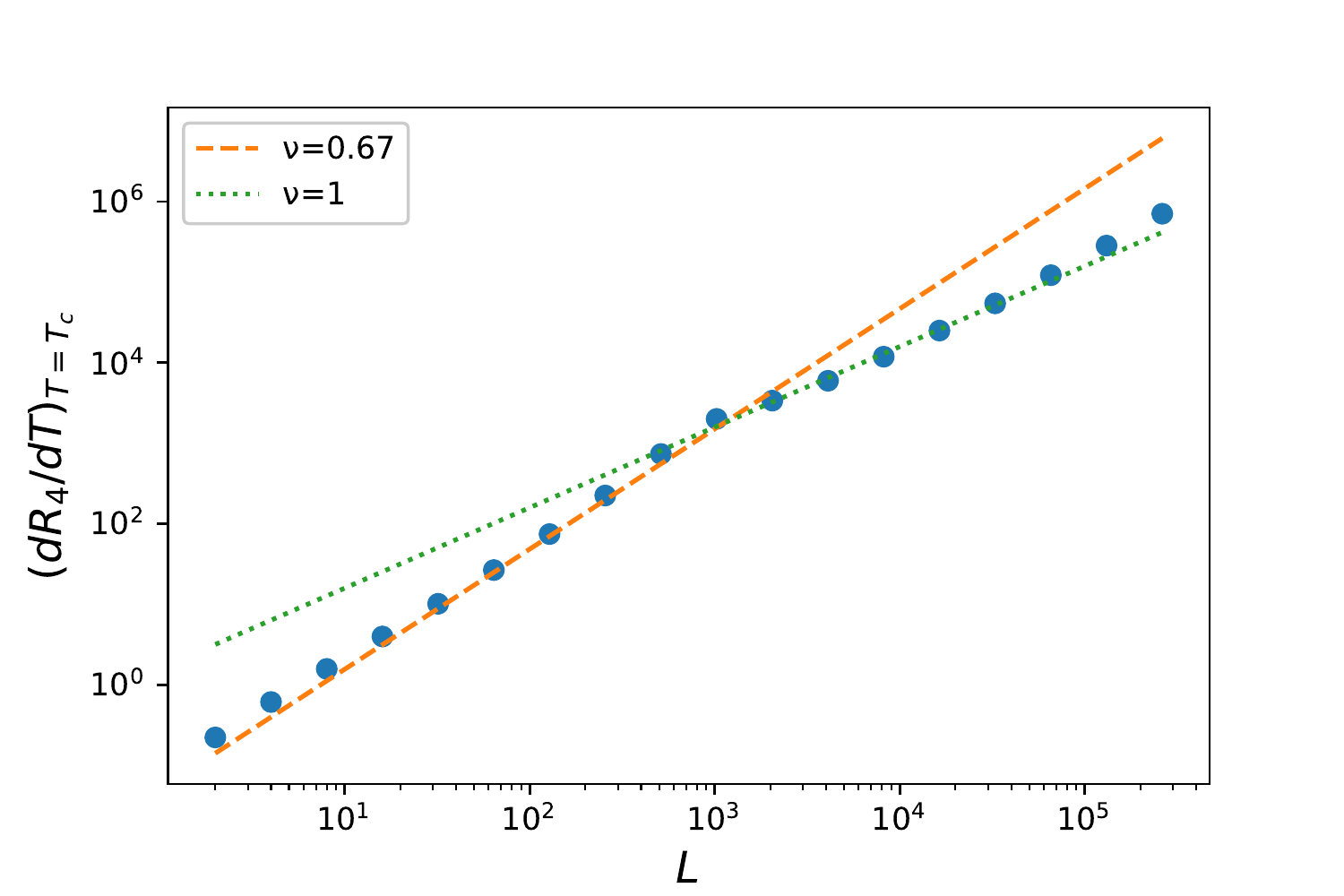}
 \caption{(Color online)System size $L$ dependence of $(dR_4/dT)|_{T=T_c}$ at $g=0.8$ with $D=32$. The dotted line represents the power law with an exponent $\nu=1.0$ for the Ising universality class, and the dashed line represents a non-trivial power law with an exponent $\nu=0.67$.}
 \label{fig:g0.8_dR4dT_FSS}
\end{figure}

Here, we discuss the universality class of the second-order phase transition for $g\geq g^*$.
At $g=0.58$, where the second-order transition behavior in $\langle E\rangle$ is observed from the result for $D=40$ in the previous subsection, $R_4$ also shows a second-order transition behavior, and the FSS of the temperature derivative of $R_4$ at $T_c$ in Eq.~(\ref{eqn:dRdT_FSS}) with $D\geq32$ yields $\nu=0.57(1)$ as shown in Fig.~\ref{fig:g0.8_dR4dT_FSS}. The evaluation of $\nu$ is based on regression using the results up to $L\leq1024$ because a numerical problem caused by a large number of renormalizations appears to exist, similar to Fig.~\ref{fig:g=0.55_dR4dT_FSS}. From the same FSS analysis of $R_4$ with different $g$, the exponent was evaluated as  $\nu=0.67(2)$ for $g=0.8$ and $\nu=0.73(2)$ for $g=1.0$. As indicated in previous studies, $\nu$ depends significantly on $g$ and increases gradually to approach $1$ with an increase in $g$. This is consistent with the fact that $\nu=1$ for the Ising universality class at $g=\infty$. However, the evaluated value of $\nu$ differs from the results obtained by MCMC\cite{Kalz2012} and transfer-matrix calculations\cite{Jin2013}.

\begin{figure}[ht]
 \centering
 \includegraphics[width=\linewidth]{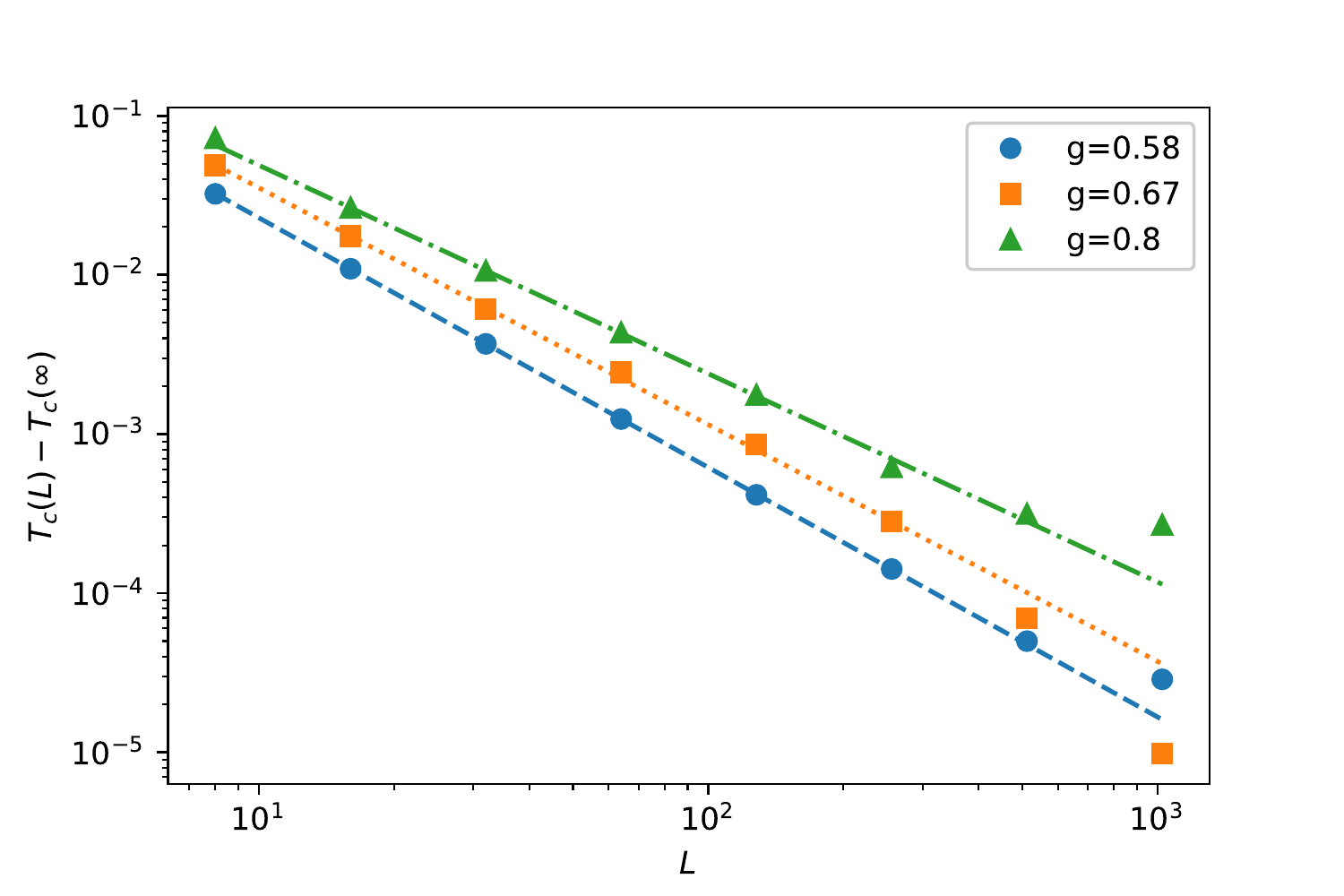}
 \caption{System size $L$ dependence of the peak temperature $T_{\text{max}}$ of the specific heat at $g=0.58$, $0.67$, and $0.8$ with $D=40$ in this HOTRG calculation. The dotted lines represent the power law $L^{1/\nu}$ with $\nu=0.638(1)$, $0.67(2)$, and $0.763(6)$, respectively.
 In addition, the critical temperatures $T_c$ are estimated to be $0.8976$, $1.1998$, and $1.5677$, respectively.}
 \label{fig:Tmax_FSS}
\end{figure}

The critical exponent $\nu$ can also be evaluated from the scaling relationship for the peak temperature of the specific heat given by Eq.~(\ref{eqn:ctemp_FSS}). Fig.~\ref{fig:Tmax_FSS} shows the results of the FSS for $g=0.58$, $0.67$, and $0.8$. Here, the exponents evaluated are $\nu=0.638(1)$, $0.67(2)$, and $0.763(5)$, respectively. The scaling for each $g$ displays a marginal deviation from the power law for large system sizes. This may be owing to the effect of the HOTRG approximation. Our evaluation of $\nu$ at $g=0.67$ and $0.8$ is in agreement with the previous MCMC studies\cite{Jin2012,Jin2013,Kalz2012}. In particular, our evaluation at $g=0.67$ (which is claimed to belong to the four-state Potts universality class in the MCMC studies) is consistent with $\nu=2/3$ (the value of the four-state Potts model).
In contrast, for $g=0.58$, which is $g^*$ in our estimation, the value of $\nu$ evaluated is significantly smaller than the lower limit of the AT model $\nu=2/3$. This is inconsistent with the previous MCMC studies\cite{Jin2012,Jin2013,Kalz2012}.

\begin{figure}[ht]
 \centering
 \includegraphics[width=\linewidth]{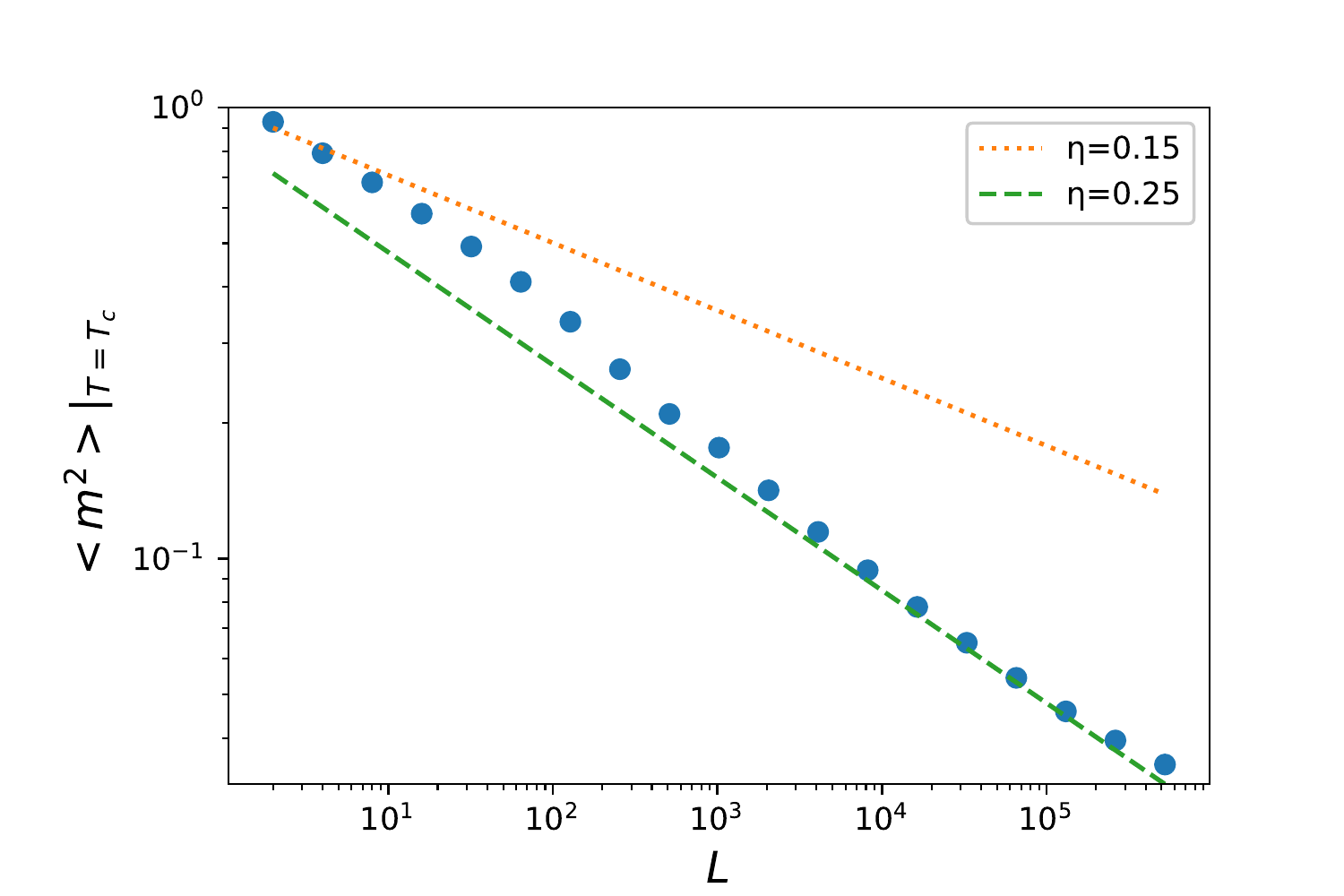}
 \caption{(Color online)System size $L$ dependence of the squared magnetization $\langle m^2 \rangle$ at $T=T_c$ and for $g=0.58$. The dotted and dashed lines represent algebraic functions with exponents $\eta=0.15$ and $0.25$, respectively. The critical temperature is estimated to be $\beta_c=1.113959$}
 \label{fig:g0.58_m2_FSS}
\end{figure}

To study the critical properties at $g=0.58$, the value of another critical exponent $\eta$ is evaluated using the FSS of Eq.~(\ref{eqn:FSS_m2}).
Fig.~\ref{fig:g0.58_m2_FSS} shows the system-size dependence of $\langle m^2\rangle$ at $g=0.58$. Here, $\eta=0.25$ is estimated from the power-law behavior at large sizes. In the previous study\cite{Li2021}, the system at $g^*$ belongs to the universality class of the tricritical Ising model. However, the value of the exponent it predicts,  $\eta=0.15$, is difficult to determine from the behavior of our results of $\langle m^2\rangle$.

\begin{figure}[ht]
 \centering
 \includegraphics[width=\linewidth]{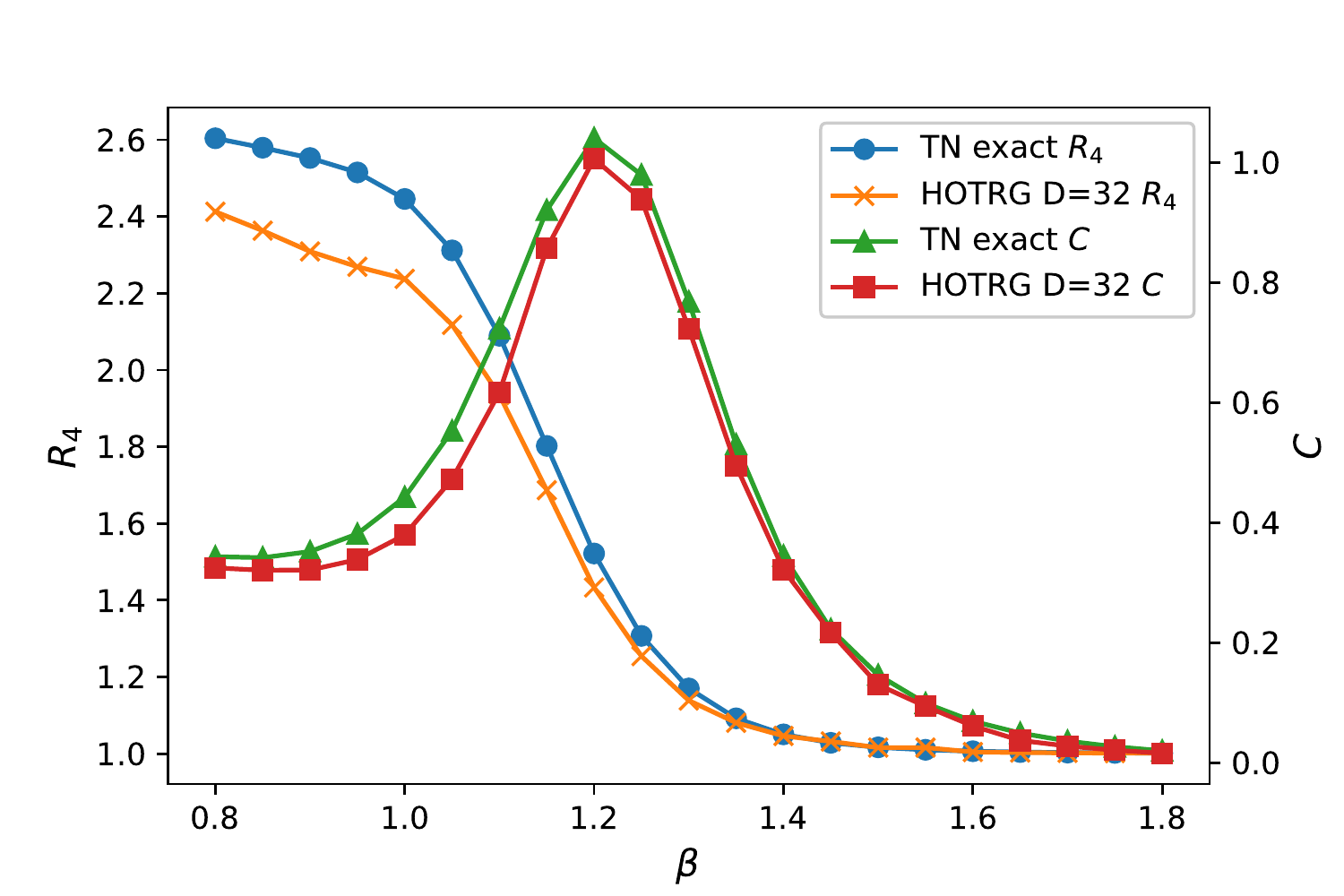}
 \caption{(Color online)
 Inverse-temperature dependence of the Binder parameter(left axis) and specific heat(right axis) of the $J_1$-$J_2$ Ising model at $g=0.55$ with $N=64$. For the Binder parameter, the circles and crosses represent exact numerical calculations and HOTRG calculations with $D=32$, respectively. For the specific heat, the triangles and squares represent exact and HOTRG calculations, respectively.
The transition temperature of this system is determined to be $\beta_c=1.2963(2)$ from the asymptotic behavior of the Binder parameter.
 }
 \label{fig:R4_C_exact}
\end{figure}

In our analysis, the difference in the evaluation of the critical exponent between the Binder parameter and specific heat can be explained by the effect of the HOTRG approximation. Fig.~\ref{fig:R4_C_exact} shows the specific heat and Binder parameter for the type-I TN at $g=0.55$ and $N=64$ obtained by the exact numerical contraction and the HOTRG calculation with $D=32$.
These results show that the specific heat is a better approximation than the Binder parameter at least for system sizes that can be calculated exactly. Therefore, the value of $\nu$ estimated from the specific heat is considered to be more reasonable than that estimated from the Binder parameter.

\begin{figure}[ht]
 \centering
 \includegraphics[width=\linewidth]{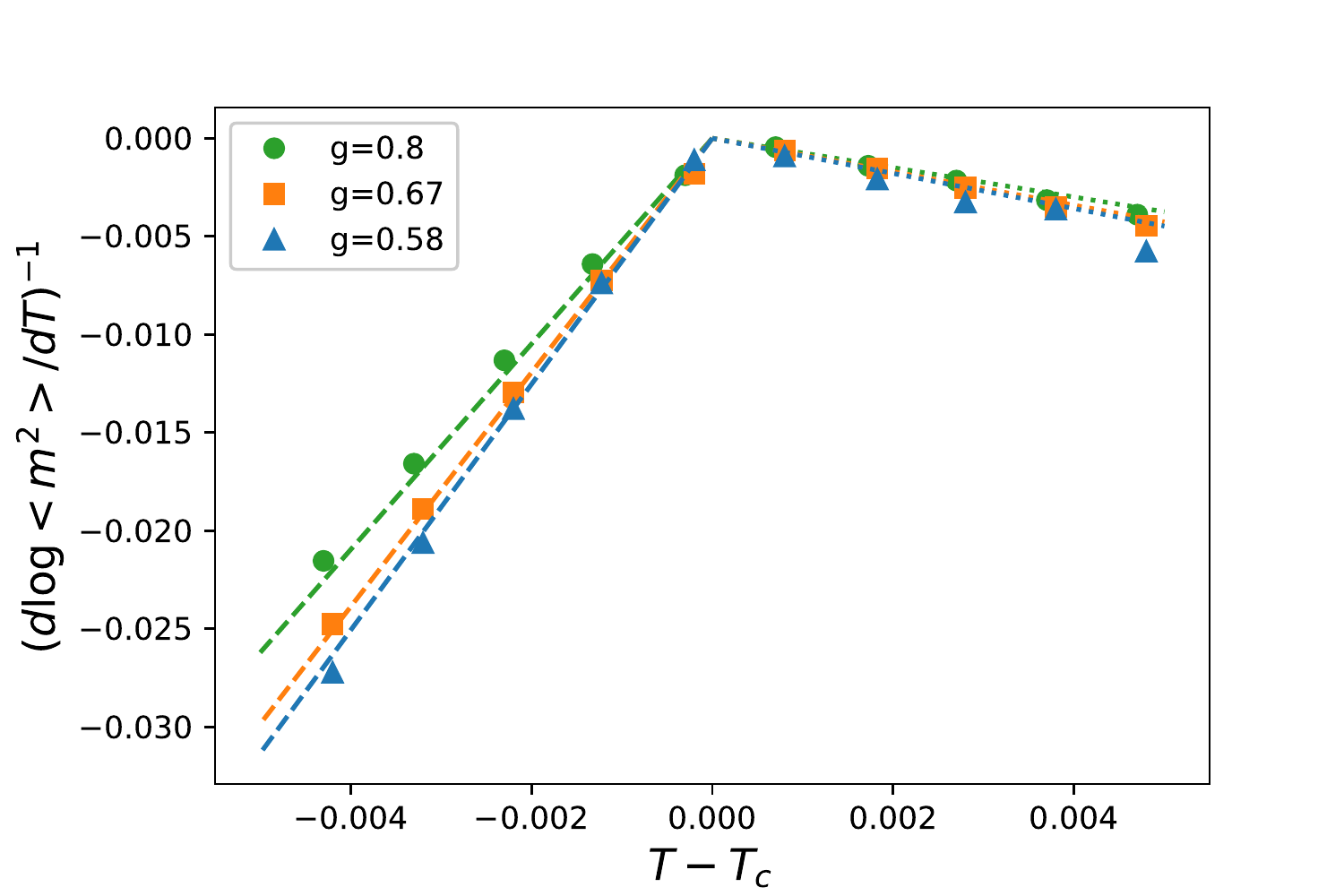}
 \caption{(Color online)
Inverse of the logarithmic derivative of the squared stripe magnetization as a function of $T-T_c$ for the $J_1$-$J_2$ Ising model  for $D=40$, $L=32768$ at $g=0.58$(triangles), $0.67$(squares), and $0.8$(circles). The dashed and dotted lines represent straight lines expected near the critical temperature whose slopes are given by the critical exponents $\beta$ and $\gamma$, depending on $g$. The values of these exponents are determined under an assumption of weak universality by using the values $\nu=0.638$, $0.67$, and $0.763$ obtained from the analysis of the specific heat for $g = 0.58, 0.67$, and $g=0.8$, respectively. The critical temperatures $T_c$ for $g=0.58$, $0.67$, and $0.8$ are estimated to be $0.8977$, $1.1997$, and $1.5678$, respectively.}
 \label{fig:dlogm_scaling}
\end{figure}

As described above, we have shown the results in which the critical exponent $\nu$ depends explicitly on $g>g^*$. We next discuss the weak universality class based on the FSS with Eq.~(\ref{eqn:logm_scaling}). As discussed in Sec.~\ref{sec:result_g<1/2} with Fig.~\ref{fig:g049_dlogM} for $g<1/2$, the behavior of the logarithmic derivative of $\langle M^2\rangle$ is a straight line with the slope of the value of critical exponents near the transition temperature.  Fig.~\ref{fig:dlogm_scaling} represents the temperature dependence of the logarithmic derivative of $\langle m^2 \rangle$ at $g=0.58$, $0.67$, and $0.8$ where the second-order transition occurs. It is evident from the figure that the slope of the straight lines depends on $g$. This indicates that the critical exponents depend explicitly on $g$, similar to $\nu$ discussed above. Furthermore, the dotted and dashed lines in the figure represent straight lines with the slope of the values of $\beta$ and $\gamma$ respectively. These exponents are estimated from the assumption of the weak universality of Eq.~(\ref{eqn:weak-universality}) and the value of $\nu$ obtained from the peak temperature of specific heat above. Although the results for $T>T_c$ (particularly at $g=0.58$) are moderately scattered, the plots and the dotted lines almost agree with each other. This indicates that the weak universality holds for $g\geq g^*$, including $g=0.58$. Here, the value of $\nu$ is estimated to be $\nu<2/3$.

\section{Discussions and Summary}
\label{sec:discussion}
First, we compare our results with those obtained by previous studies. The previous study by the iTEBD method\cite{Li2021}, a tensor network method, using the type-I TN described in this work observed that for $g=0.55$, $\langle E\rangle$ and $\langle m^2\rangle$ vary continuously without jumps at the transition temperature. It concluded that the transition is a second-order transition. In our HOTRG calculations using the type-I TN, we also observed that $\langle E\rangle$ varies continuously when the bond dimension $D$ is small(see Fig.~\ref{fig:g=0.55_E_Ddep}). However, it is verified that $\langle E\rangle$ varies discontinuously when $D$ is large. Our results imply the first-order transition in the limit of $D\to\infty$. This indicates the need to reexamine the conclusion of the previous study using the iTEBD method for the effect of a finite $D$.

Meanwhile, as shown in Fig.~\ref{fig:g0.575_E}(b), $\langle E\rangle$ at $g=0.58$ behaves similarly to the first-order transition for a small $D$. However, it transforms to behave similarly to the second-order transition as $D$ increases. Similarly, for $R_4$, it is evident that an increase in $D$ at $g=0.55$ yields alternating first-order/second-order transition-like behavior. Thus, the behavior of the physical quantities at a finite $D$ is rather complex. Even when a first-order transition-like result is obtained for a relatively large $D=40$, the possibility of an eventual second-order transition at $D\to\infty$ cannot be precluded completely.

The effect of a finite $D$ in HOTRG is an issue to be considered. Meanwhile, the capability to compute larger sizes compared with MCMC is an advantage. Our HOTRG calculations indicate that the boundary between the first-order and second-order transitions is $g^*\simeq 0.58$. This is smaller than $g^*=0.67(1)$ in the previous study by MCMC\cite{Jin2012}. In particular, our results for $g>0.62$ show the second-order transition behavior independent of the value of $D$. This strongly indicates that the region of the first-order transition, if it exists, is narrower than that estimated by the MCMC study.
In contrast, we also consider that a finite region of the first-order transition exists. For example, we verify that at $g=0.55$, the thermodynamic relationship that should hold at the phase boundary of the first-order transition is satisfied. The details are provided in Appendix~\ref{sec:C.C.relation}.

With regard to the estimation of $g^*$, our results differ from the previous results on the issue of universality class as well. The previous MCMC study\cite{Jin2013,Kalz2012} asserted that the critical property for $g>g^*$ belongs to the AT universality classes and that the universality class of the four-state Potts model, namely, $\nu=2/3$, holds at the endpoint $g=g^*$. Our results show that the weak universality holds as anticipated from the AT universality class. Furthermore, the critical exponents vary continuously with $g$ while maintaining the ratio of exponents. This is consistent with the previous study. However, corresponding to the extension of the region of $g$ for the second-order transition that we evaluated, it is indicated that the value of $\nu$ may be significantly smaller than that of the four-state Potts model. This is inconsistent with the AT scenario where the entire domain of $g>g^*$ of the $J_1$-$J_2$ Ising model is mapped to the AT universality class. This result indicates that the critical behavior of this model is closer to the eight-vertex model\cite{Baxter2016} that varies $\nu>1/2$ under a similar weak universality class than to the AT model that adopts $\nu>2/3$, although the microscopic correspondence is ambiguous.

Next, we discuss the numerical accuracy of the renormalization process in HOTRG calculations observed in this study. It is shown in practice that certain physical quantities of interest are affected by the approximation in HOTRG calculations as the number of renormalization steps (i.e., the system size) increases.
In the size dependence of $(\frac{dR_4}{dT})|_{T=T_c}$ for $g=0.8$ shown in Fig.~\ref{fig:g0.8_dR4dT_FSS}, although it follows a power law with the exponent $\nu\simeq 0.67$ up to approximately $L\leq1024$, a crossover behavior to another power law with $\nu= 1$ is observed for $L\geq 1024$. This crossover behavior is also observed for $g=0.55$, as shown in Fig.~\ref{fig:g=0.55_dR4dT_FSS}.
In addition, Fig.~\ref{fig:g=0.55_C} shows the crossover observed in the specific heat, where a strong divergence trend is observed for the relatively small $D=28$ at a small $L$, whereas the divergence trend weakens dramatically as $L$ increases. This crossover in size dependence disappears as $D$ increases. Therefore, this can be considered to be caused by the approximations of HOTRG owing to the small $D$.
These size dependencies may then be interpreted as a pseudo-appearance of the Ising universality as large sizes because these are explained by the exponent $\nu=1$ and $\alpha=0$. This interpretation is also compatible with the fact that as shown in Fig.~\ref{fig:g1.0_X}, a plateau of $X=2$ is observed for a sufficiently large $L$ immediately below the transition temperature. That is, the degree of freedom in the ordered phase appears to be two, similar to the Ising model.
To prevent the influence of the pseudo-behavior in the estimation of the critical exponents, we use the sizes in our FSS analysis up to the appearance of the power law of the Ising universality at large sizes in Sec.~\ref{sec:result_g<g^*} and Sec.~\ref{sec:result_g>g^*}.

If the above argument is correct, it can be determined that in the renormalization procedure of HOTRG, the four-hold symmetry expected from the low-temperature phase for $g>1/2$ is missing in the two-hold symmetry owing to the approximation. A possible reason is that the HOTRG renormalization procedure breaks the $x$-$y$ symmetry of the square lattice. Because the TRG method\cite{Levin2007} tilts the lattice by $\pi/4$ unlike HOTRG, the renormalization procedure does not explicitly depend on the $x$- and $y$-directions. Thereby, the symmetry in the $x$-$y$ direction may be preserved.

\begin{figure}[ht]
 \centering
 \includegraphics[width=\linewidth]{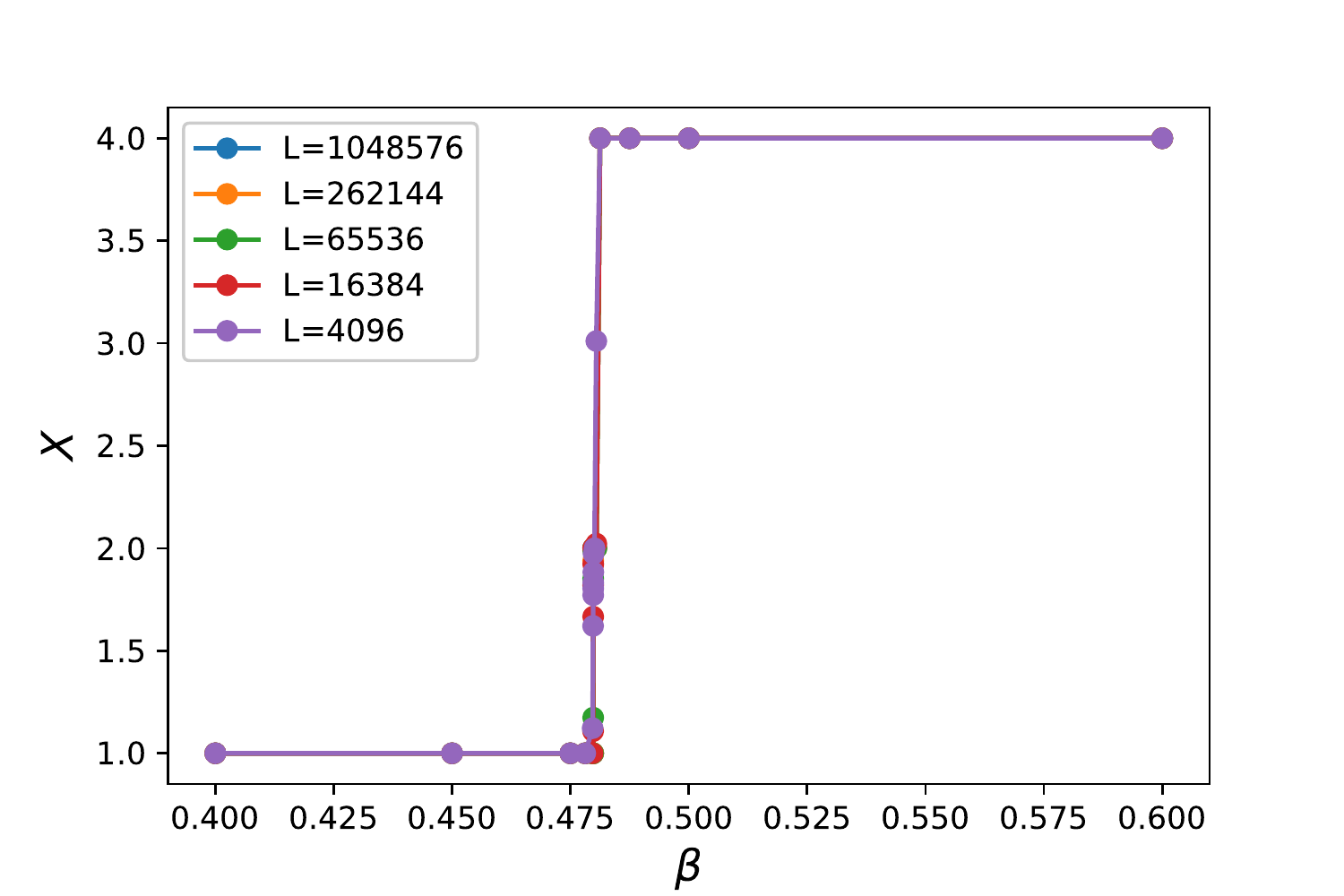}
 \caption{(Color online)Inverse-temperature dependence of $X$ of type-II TN for the $J_1$-$J_2$ Ising model at $g=1.0$ and with $D=32$ by the TRG method.}
 \label{fig:X_TRG}
\end{figure}

We calculated $X$ using the TRG method with two tensor networks: type-I, and type-II. As shown in Fig.~\ref{fig:X_TRG}, a direct transition from $X=1$ to $4$ with almost no through the intermediate state of $X=2$ only in the case of the TRG method combined with the type-II TN.
In general, under conditions fixed to the same $D$, the numerical accuracy of physical quantities is better for HOTRG than for TRG\cite{Xie2012}, and for the type-I TN than for the type-II TN (as discussed in Appendix~\ref{sec:TN_selection}). However, it should be noted that the higher accuracy of physical quantities does not necessarily imply that of $X$. One should also consider the symmetry of the tensors used, etc, while studying the properties of a renormalized tensor such as $X$.

\begin{figure}[ht]
 \centering
 \includegraphics[width=\linewidth]{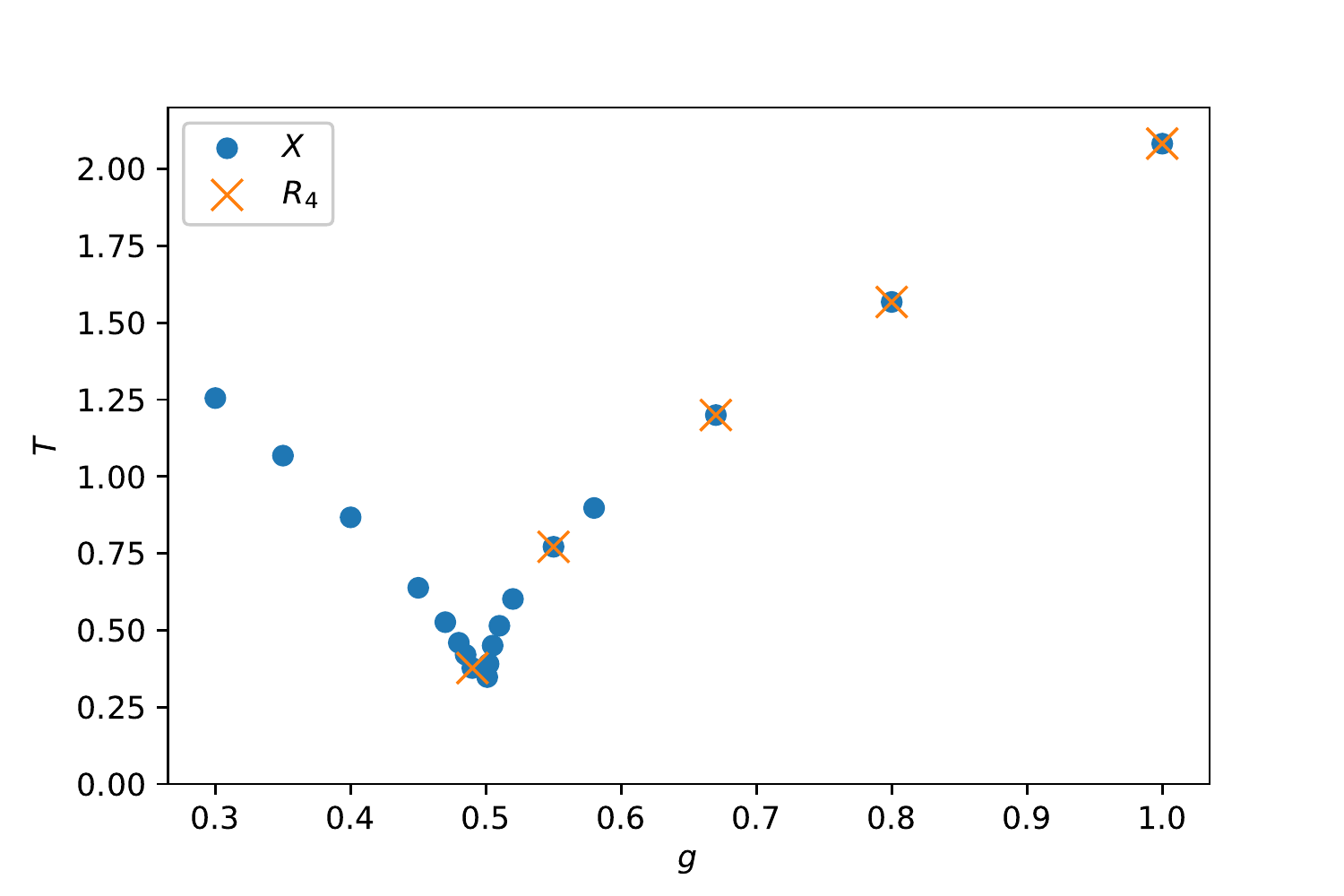}
 \caption{(Color online)Phase diagram of the $J_1$-$J_2$ Ising model in the plane of temperature $T$ and coupling constant $g$. The circles represent the transition temperature at which $X$ in HOTRG with $D=32$ jumps from $1$ to $2$ for each $g$, and the crosses represent the transition temperature obtained by the Binder parameter $R_4$ for sufficiently large sizes. }
 \label{fig:phase_diagram_HOTRG}
\end{figure}

Although the symmetry missing in the HOTRG method occurs in the low-temperature phase, the quantity $X$ still displays important properties of the renormalized tensor. Furthermore, the phase diagram can be obtained from the boundary at which $X=1$ is unstable. As has been indicated, $X$ is a quantity evaluated from a renormalized tensor by the method of the general tensor renormalization groups. Measuring this quantity has the advantage that it can be calculated as a by-product without the need to calculate physical quantities using the impurity tensor method. Certain systems have already used this $X$ to evaluate transition temperatures\cite{Li2022}\cite{Jha2022}. Here, the phase boundary obtained as the temperature at which $X$ jumps from $1$ to $2$ in HOTRG is shown in Fig~\ref{fig:phase_diagram_HOTRG}, in conjunction with the transition temperature obtained by the Binder parameter. It is observed that the transition temperatures evaluated by the two methods coincide with each other.

To summarize, we studied the critical phenomena of the $J_1$-$J_2$ Ising model by varying the parameter $g$ using HOTRG. Our results for $g<1/2$ indicate that the critical properties of the second-order phase transition are explained by the universality class of the two-dimensional Ising model. This is in agreement with previous studies. For $g>1/2$, although the influence of a finite $D$ should be paid due attention, various numerical results such as the jump in internal energy at $T_c$, $L^2$ divergence of specific heat, and a sharp peak of the Binder parameter strongly indicate the existence of the first-order transition region near $g=1/2$. We estimated the upper boundary of the region to be $g^*\simeq 0.58$. This value is smaller than the previous MCMC result, $g^*=0.67(1)$\cite{Jin2012}. This results in a narrower region of first-order transition (if any) than concluded in the previous study.

For the universality class of the second-order phase transition for $g>g^*$, our results support the assertion that the weak universality holds for any $g$. This is consistent with the previous MCMC study\cite{Jin2013,Kalz2012}, but incompatible with the result supporting the tricritical Ising universality class with $\gamma/\nu=37/20$. Meanwhile, our results do not fully support the AT scenario. To be specific, the value of $\nu$ we obtained at $g=0.58$ with the second-order phase transition is significantly smaller than $2/3$. This indicates that the universality class of the eight-vertex model\cite{Sutherland1970,Baxter1971,KadanoffWegner1971,Baxter2016} with the same weak universality may be valid, rather than the AT universality class.

The eight-vertex model can adopt a value of $\nu$ larger than $1/2$. Therefore, it may be reasonable to adopt values smaller than our evaluated value of $\nu=0.638$ at $g=g^*$. Thus, it is still feasible to exhibit a second-order phase transition to a region closer to $g=1/2$ than $g^*=0.58$ obtained in this study. Considering this, a more accurate determination of the location of the critical endpoint $g^*$ would be undertaken in future work in conjunction with an improvement of the accuracy of the tensor renormalization group methods.

Finally, we discuss the tensor renormalization group methods from a methodological perspective. Although the system displays a four-fold symmetry for $g>1/2$ in the $J_1$-$J_2$ Ising model,  the HOTRG calculations show that this four-fold symmetry is missing at a certain stage of the renormalization for certain physical quantities. Presumably, this is the reason why the critical behavior of certain quantities is Ising-like, which reflects the two-fold symmetry after the symmetry is missing. It is also verified that such four-fold symmetry may be preserved by a TRG method rather than HOTRG. The capability to calculate using TRG-like methods such as bond-weighted TRG\cite{Adachi2022} to preserve the symmetry of the system up to larger system sizes exhibits a high potential.

\appendix
\section{numerical validity evaluation}
\subsection{Tensor network construction methods and numerical accuracy}
\label{sec:TN_selection}
\begin{figure}[ht]
 \centering
 \includegraphics[width=\linewidth]{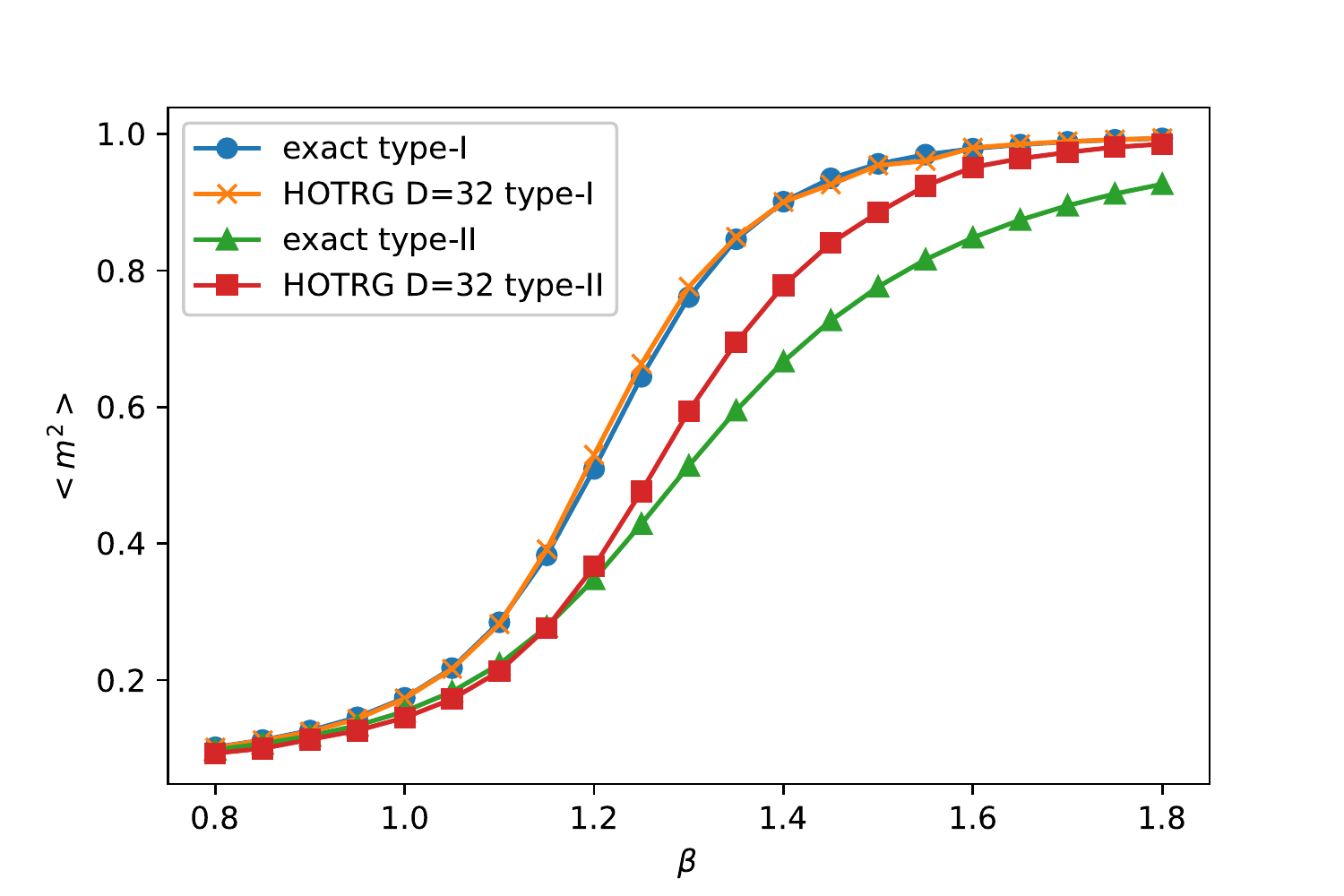}
 \caption{(Color online)Inverse-temperature dependence of the squared order parameter $\langle m^2\rangle$ of the system at $g=0.55$ obtained by HOTRG with the type-I TN by defined Eq.~(\ref{eqn:TN1_a}) and (\ref{eqn:TN1_b})(circles), its exact calculations(crosses), the type-II TN by Eq.~(\ref{eqn:TN2}) (squares), and its exact calculations(crosses)}
 \label{fig:TN:I-II_M2}
\end{figure}

In general, there are several tensor network (TN) representations for a system. In Sec.~\ref{sec:methods}, we introduce two specific representations for the $J_1$-$J_2$ Ising model, (see Fig.~\ref{fig:TNa,b}): the type-I TN defined by Eq.~(\ref{eqn:TN1_a}) and (\ref{eqn:TN1_b}), and the type-II TN defined by Eq.~(\ref{eqn:TN2}).
For the system with $N=64$, we perform HOTRG calculations with $D=32$ fixed for each of the two representations, in addition to the exact calculations in each representation. Fig.~\ref{fig:TN:I-II_M2} presents the numerical results of the squared order parameters as a function of the inverse temperature at $g=0.55$. The results of the type-I TN are in agreement with the exact calculations at all temperatures. However the type-II TN displays large errors, particularly at low temperatures. This indicates that the type-I TN of Eq.~(\ref{eqn:TN2}) is significantly more accurate.

\begin{figure}[ht]
 \centering
 \includegraphics[width=\linewidth]{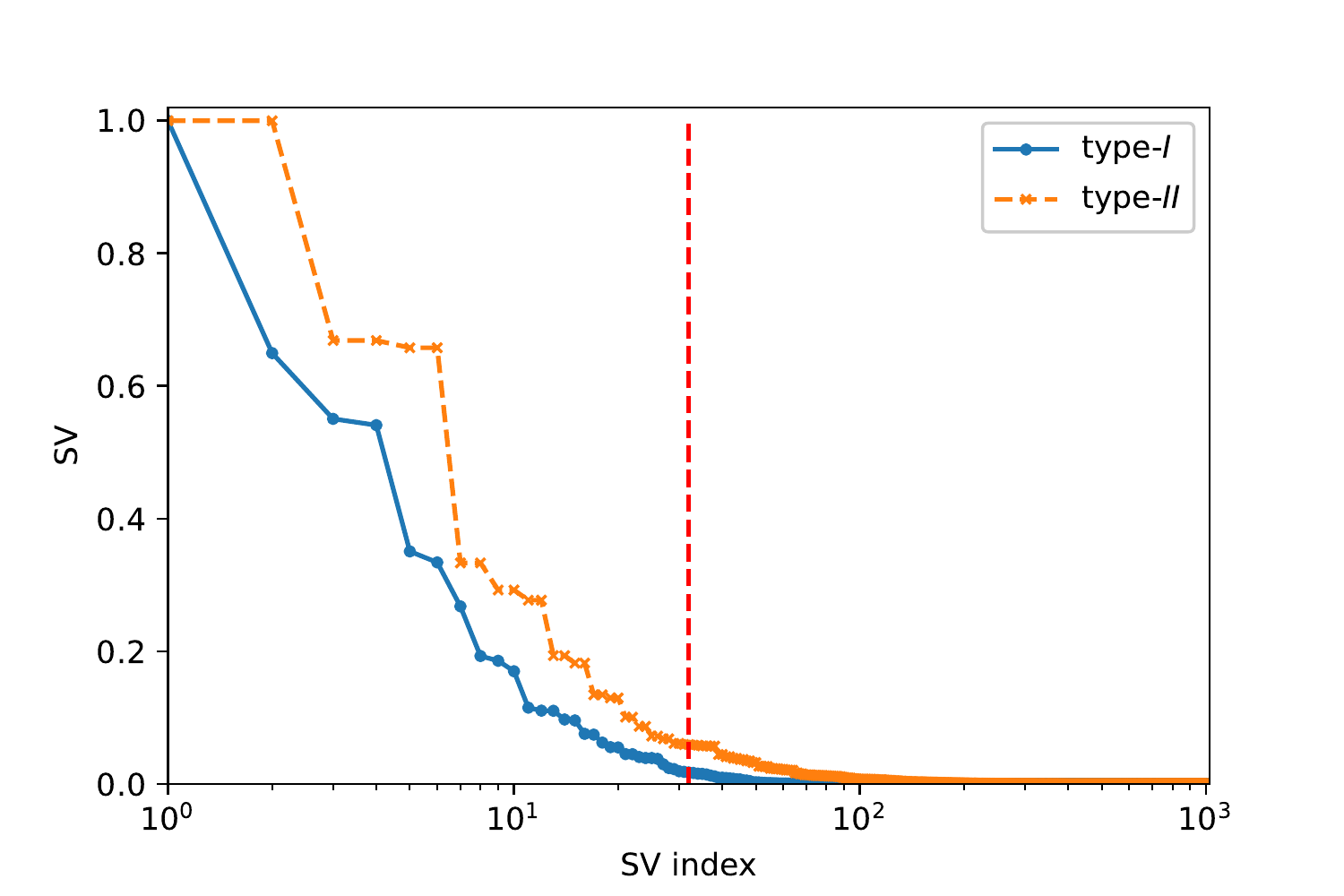}
 \caption{(Color online)Index dependence of the singular value for the type-I TN(solid) and the type-II TN(dashed) at the third renormalization step in HOTRG with $D=32$ of the system for $g=0.55$ at $\beta=1.3$ which is lower than the transition temperature. The vertical line indicates an index of $32$.}
 \label{fig:SV}
\end{figure}

We discuss this result from the perspective of the distribution of singular values at the approximations in the renormalization steps. Fig.~\ref{fig:SV} shows the distribution of singular values at the third renormalization step where the approximation procedure first appears in the HOTRG with $D=32$. The vertical dotted line represents the index of the singular value with $D=32$, and the renormalization step discards the singular values on the right side of the line while retaining those on the left side. Therefore, the accuracy of the approximation improves with the smaller area on the right side relative to that on the left side of the dotted line. Thus, it is evident from the figure that the type-I TN is more accurate than the type-II TN.

The reason for this slow convergence of the distribution of the singular values of the type-II TN is that the singular values are degenerate. This may be because in the type-II TN setup, all the spin states are included in the two tensor indices, which results in a redundant representation. Therefore, it is recommended that such redundant TN settings be avoided in general.

\subsection{Clausius--Clapeyron relation under a uniform magnetic field}
\label{sec:C.C.relation}
\begin{figure}[ht]
 \centering
 \includegraphics[width=\linewidth]{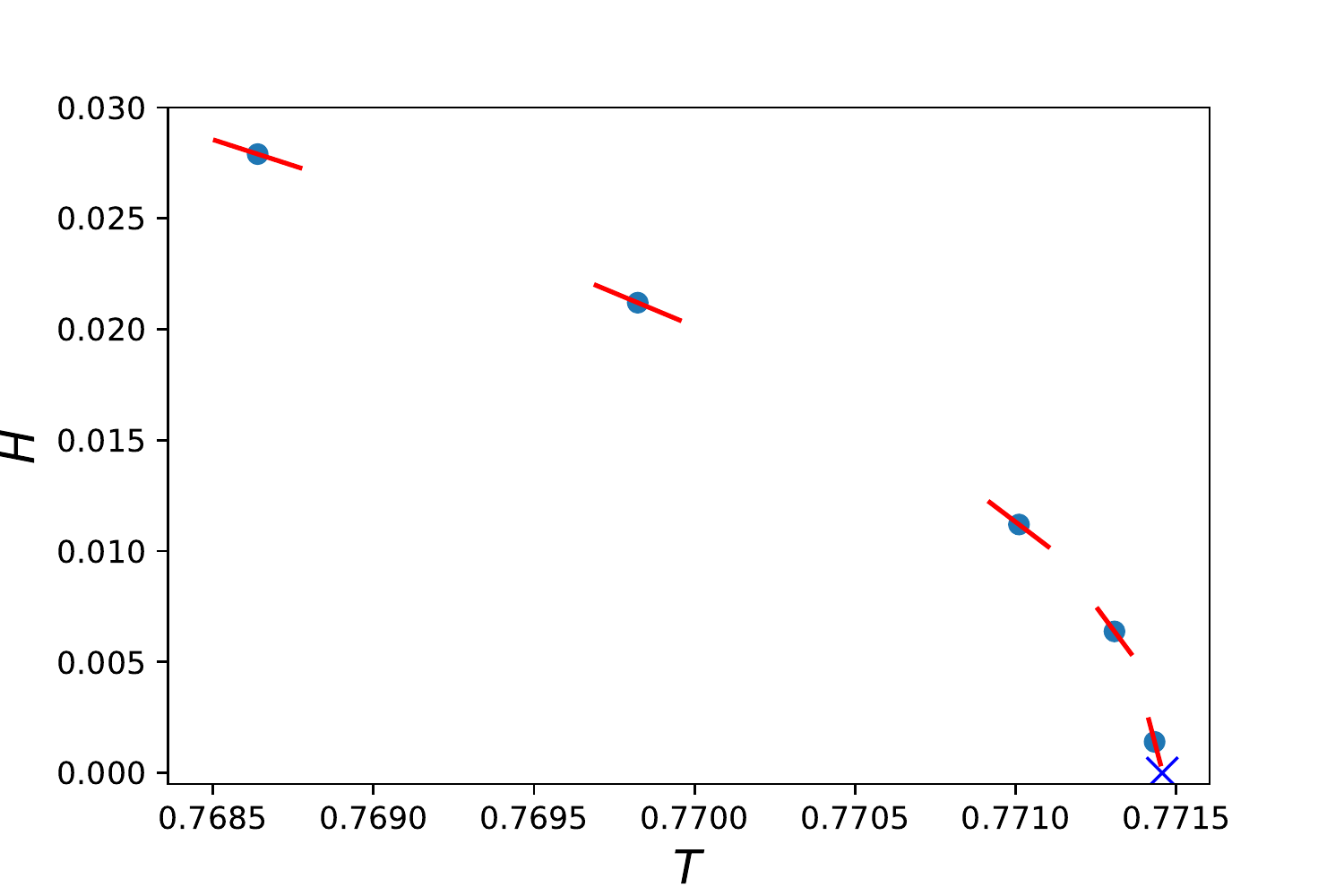}
 \caption{(Color online)Phase diagram of the $J_1$-$J_2$ Ising model at $g=0.55$ in the plane of temperature $T$ and uniform magnetic field $H$. The circles indicate the transition temperatures at which the magnetization jumps (estimated by HOTRG). The lines represent the slope of the phase boundary evaluated from the right-hand side of the Clausius--Clapeyron relation (Eq.~(\ref{eqn:C.C})). The cross at $H=0$ represents the first-order transition temperature of the stripe order parameter $\langle m^2\rangle$. }
 \label{fig:phase_diagram_H}
\end{figure}

When a system exhibits a first-order phase transition, a consequence of equilibrium thermodynamics is that its coexistence curve satisfies the Clausius--Clapeyron relation\cite{Fermi}. In numerical calculations, this relation should be satisfied if a first-order transition actually occurs. We investigated this relation as an additional supporting evidence for the first-order transition. Consider the phase diagram of a general magnetic system in the plane of a uniform magnetic field $H$ and temperature $T$, and let the coexistence curve be $H_c(T)$. The stripe order realized at a low temperature in the $J_1$-$J_2$ Ising model for $g>1/2$ discussed in this study is orthogonal to the uniform field. Therefore, the stripe phase is likely to be stable under the field.
In this case, the Clausius--Clapeyron relation is given by
\begin{equation}
    \frac{dH_c}{dT} = \frac{1}{T_c}\left(\frac{U_A-U_B}{M_B-M_A} -H_c\right),
    \label{eqn:C.C}
\end{equation}
where for the coexistent phases $A$ and $B$, the internal energy $U_A$ and $U_B$, and the uniform magnetization $M_A$ and $M_B$ are defined as
\begin{equation}
  \begin{cases}
    \displaystyle U_A  = \lim_{H\to H_c-0}U,\\
    \displaystyle U_B  = \lim_{H\to H_c+0}U,\\
    \displaystyle M_A  = \lim_{H\to H_c-0}M,\\
    \displaystyle M_B  = \lim_{H\to H_c+0}M
  \end{cases}
\end{equation}
, respectively.
Fig.~\ref{fig:phase_diagram_H} shows the phase diagram of the $J_1$-$J_2$ Ising model under the uniform magnetic field at $g=0.55$.
The transition field $H_c$ of the first-order transition and the physical quantities in each phase $U_A$,$U_b$, $M_a$, and $M_B$ were estimated by HOTRG calculation with $D=32$ while varying the magnetic field with a fixed temperature.
The slope of the phase boundary (calculated from the right-hand side of Eq.~(\ref{eqn:C.C})) is drawn as lines on the points of each transition field. This is consistent with the phase boundary profile.
Furthermore, the transition temperature value in the $H_c\to 0$ limit in the phase diagram approximately corresponds to the first-order transition temperature where the stripe magnetization jumps at $H=0$. The slope of the phase boundary appears infinite in the limit. This is consistent with the fact that at $H=0$. The internal energy has a finite jump and the uniform magnetization is continuous at the transition temperature.
Thus, it is verified that the thermodynamic relation in the case of the first-order phase transition is satisfied.

\begin{acknowledgments}
This work was supported by MEXT as the Program for Promoting Research on the Supercomputer Fugaku (DPMSD, Project ID: JPMXP1020200307). One of the authors, KY, was supported by the SPRING-GX program at the University of Tokyo.
\end{acknowledgments}

\bibliographystyle{apsrev4-2}
\bibliography{J1J2_paper}
\end{document}